\documentclass{article}

\usepackage{PRIMEarxiv}

\usepackage{amsmath,amsfonts,bm}

\def\eqref#1{equation~\ref{#1}}

\def\1{\bm{1}}

\def\rvg{{\mathbf{g}}}

\def\rvn{{\mathbf{n}}}

\def\rvr{{\mathbf{r}}}

\def\rvx{{\mathbf{x}}}
\def\rvy{{\mathbf{y}}}

\DeclareMathAlphabet{\mathsfit}{\encodingdefault}{\sfdefault}{m}{sl}
\SetMathAlphabet{\mathsfit}{bold}{\encodingdefault}{\sfdefault}{bx}{n}

\def\gG{{\mathcal{G}}}

\def\gR{{\mathcal{R}}}

\usepackage{url}
\usepackage{graphicx}
\usepackage[export]{adjustbox}
\usepackage{subcaption}
\usepackage{booktabs}
\usepackage{multirow}
\usepackage{placeins}
\usepackage{bm}
\usepackage{enumitem}
\usepackage{array}
\newcommand{\aref}[1]{\hyperref[#1]{Appendix~\ref*{#1}}}

\usepackage[some]{background}
\backgroundsetup{%
scale=1,  
angle=0,  
opacity=0.7,  
color = {darkgray},  
contents={\parbox{\textwidth}{\centering 
\large This paper has been accepted for publication in \\ Transactions on Machine Learning Research \\[26cm]}
}
}

\usepackage[utf8]{inputenc} 
\usepackage[T1]{fontenc}    
\usepackage{url}            
\usepackage{booktabs}       
\usepackage{amsfonts}       
\usepackage{nicefrac}       
\usepackage{microtype}      
\usepackage{lipsum}
\usepackage{natbib}
\usepackage{fancyhdr}       
\usepackage{graphicx}       
\graphicspath{{media/}}     
\usepackage[colorlinks = true,
            linkcolor = blue,
            urlcolor  = blue,
            citecolor = blue]{hyperref} 

\title{Unifi3D: A Study on 3D Representations for Generation and Reconstruction in a Common Framework
}

\author{
 Nina Wiedemann${}^{*}$, Sainan Liu${}^{*}$, Quentin Leboutet${}^{*}$, Katelyn Gao${}^{}$, Benjamin Ummenhofer, \\ \vspace{-3pt} \\ \textbf{Michael Paulitsch, Kai Yuan} \\
  \vspace{1em}\\ Intel Corporation
}

\begin{document}
\maketitle
\BgThispage

\begingroup
\renewcommand\thefootnote{\textasteriskcentered}
\footnotetext{Equal contribution. Correspondence: \texttt{nina.wiedemann@intel.com}}
\endgroup

\begin{abstract}
Following rapid advancements in text and image generation, research has increasingly shifted towards 3D generation. Unlike the well-established pixel-based representation in images, 3D representations remain diverse and fragmented, encompassing a wide variety of approaches such as voxel grids, neural radiance fields, signed distance functions, point clouds, or octrees, each offering distinct advantages and limitations. 
In this work, we present a unified evaluation framework designed to assess the performance of 3D representations in reconstruction and generation. We compare these representations based on multiple criteria: quality, computational efficiency, and generalization performance. Beyond standard model benchmarking, our experiments aim to derive best practices over all steps involved in the 3D generation pipeline, including preprocessing, mesh reconstruction, compression with autoencoders, and generation. Our findings highlight that reconstruction errors significantly impact overall performance, underscoring the need to evaluate generation and reconstruction \textit{jointly}. 
We provide insights that can inform the selection of suitable 3D models for various applications, facilitating the development of more robust and application-specific solutions in 3D generation.
The code for our framework is available at \url{https://github.com/isl-org/unifi3d}.
\end{abstract}


\section{Introduction}\label{sec:intro}

Recent advancements in generative image synthesis architectures, such as Generative Adversarial Networks (GANs) and Diffusion Models, have driven significant progress in the field of 3D generation~\citep{gezawa2020review, li2024advances, zhao2024challenges, liu2024comprehensive, jiang2024survey}. While image generation has reached a stage of maturity, mainly standardizing around pixel-based representations~\citep{crowson2024scalable}, the landscape for 3D representations remains fragmented and varied. A wide range of 3D representations, including Voxel Grids~\citep{ren2024xcube}, Neural Radiance Fields (NeRFs)~\citep{mildenhall2021nerf}, Signed Distance functions (SDFs)~\citep{park2019deepsdf}, Point Clouds (PC)~\citep{nichol2022point}, and Octrees~\citep{wang2022dual}, have been proposed, each suited to different applications and tasks. These methods vary not only in the way they encode geometry but also in how they handle the trade-offs between quality, computational efficiency, memory requirements, and the ability to generalize to novel objects and scenes~\citep{liu2024comprehensive, peng2020convolutional, wangdiffusion}. 

While recent advances in 3D generative models can be attributed to various factors — such as the availability of improved datasets~\citep{deitke2023objaverse}, optimized sampling techniques for diffusion models~\citep{ren2024xcube}, and enhanced generative architectures~\citep{zhang2024clay} — the choice of the underlying representation remains a key factor. 
It dictates the information loss prior to encoding, influences the models used for compression and generation, and defines the computational resources for reconstructing a mesh from a generated sample. Therefore, assessing the suitability of different 3D representations for reconstruction and generation is of paramount importance. 

\begin{figure*}[t]
    \centering
    \includegraphics[width=1\linewidth]{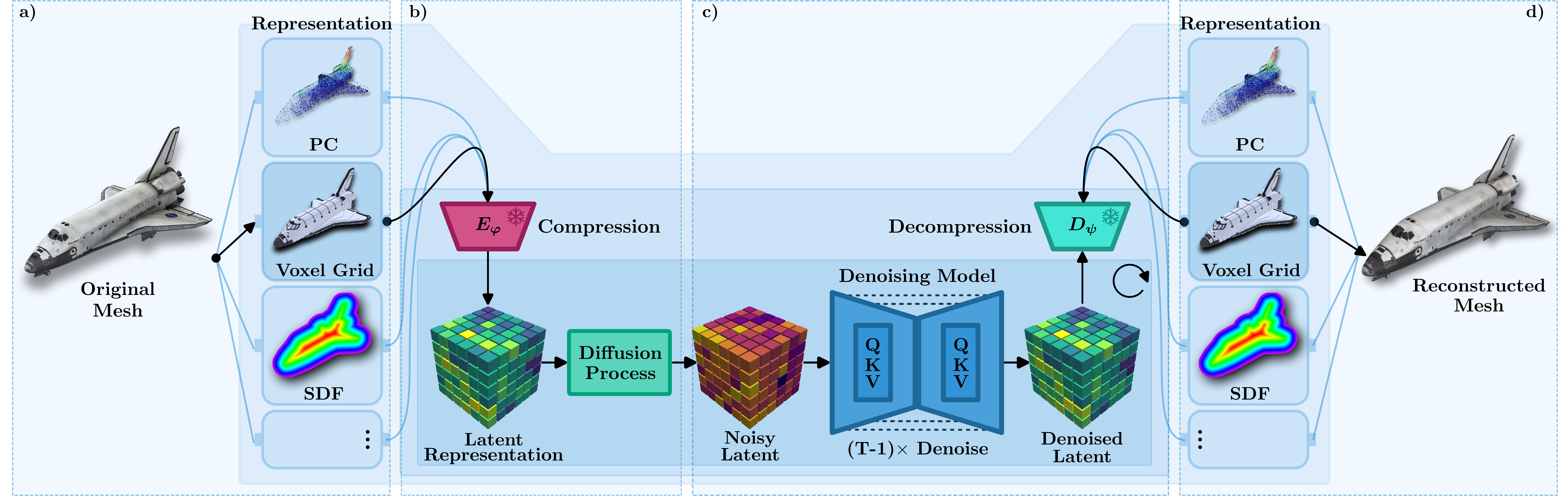}
    \caption{Overview of the steps involved in a standard 3D generation pipeline: a) the mesh is transformed into a suitable representation, b) an encoder $\boldsymbol{E_\varphi}$ is pre-trained to compress it into a latent vector, c) a diffusion model is then trained to denoise the latent, d) the latent is finally decompressed into a target representation using a pre-trained decoder $\boldsymbol{D_\psi}$ and turned back into a mesh using a dedicated algorithm such as marching-cubes. (PC: point cloud, SDF: signed distance field)}
    \label{fig:pipeline}
\end{figure*}

However, objectively comparing 3D representations based on existing literature poses a significant challenge. While the representation is a core component, it is deeply embedded within complex 3D generative pipelines that utilize various models, loss functions, and datasets. On top of that, a wide variety of pre- and post-processing techniques is applied, significantly affecting the results. 
For instance, handling non-watertight meshes is often not well-documented, which tends to create inconsistencies across studies~\citep{zhang2024clay}. 
Another key issue lies in the difficulty of evaluating 3D object quality. Traditional metrics such as Chamfer Distance (CD) are commonly used to assess geometric accuracy, but they fall short in capturing perceptual quality and finer details~\citep{mescheder2019occupancy, wu2020pq}. 
In turn, user studies can provide valuable qualitative insights, but they are labor-intensive and time-consuming. As a result, many papers resort to presenting qualitative results or cherry-picking favorable examples, which undermines objective evaluation and hinders progress in the field~\citep{zhao2024challenges}. 

This paper introduces a unified evaluation protocol designed to benchmark 3D representations. We have developed a standardized pipeline that integrates all components of the generative process — from data preprocessing to mesh encoding, generation and mesh reconstruction — into a common framework, as illustrated in \autoref{fig:pipeline}. This design allows for the interchangeable use of 3D representations, ensuring that any observed performance differences are inherent to the representations themselves. 
Our standardized pipeline ensures a fair comparison between different representations with respect to reconstruction and generation, instead of relying on the individual implementations of different 3D generation approaches that leverage the representations in varying manner, evaluate them with varying metrics, and oftentimes solely report conditional generation performance of the complete pipeline. 

In contrast to \textit{model} benchmarking, we aim to generate insights into the strengths and weaknesses of each representation while controlling for confounding factors like the diffusion model. Unlike traditional review papers that primarily offer qualitative overviews of existing methods~\citep{po2024state, wangdiffusion, cao2024survey, li2024advances, gezawa2020review, liu2024comprehensive}, our work emphasizes quantitative assessments grounded in empirical evidence.
We make our entire pipeline open source to ensure the repeatability of our results and their generalization to different representations or experimental conditions. This allows rapid prototyping of novel 3D generative methods or 3D representations while ensuring adherence to best practices, such as proper preprocessing and hyper-parameter optimization. Our codebase also introduces previously unavailable open-source components, including a novel Dual-Octrees-based generative approach, training code for Shap-E, and transformer-based occupancy network generation. 
Summarized, the contributions of this paper are:
\begin{itemize}
    \item \textbf{Standardized generation pipeline:} We implement a generation pipeline with plug-and-play functionality to test 3D representations (voxel grids, SDFs, point clouds, octrees, triplanes, and NeRFs) in a latent diffusion setting as the most common approach for 3D generation. 
    \item \textbf{Jointly evaluating reconstruction and generation:} We analyze the relation between reconstruction and generation capability. Reconstruction errors are as high as 20\% of the generation error, proving their significance.
    \item \textbf{Best practices:} 
    We provide insights for common problems and derive best practices, such as the effects of different data preprocessing methods and the choice of meaningful sample sizes for evaluation.
    \item \textbf{Open-Source Modular Codebase:} We provide an open-source codebase with a structured, modular architecture (see \autoref{fig:pipeline}) for rapid development and evaluation of new methods with various generation models and 3D representations. 
\end{itemize}

\section{Benchmarking 3D Representations and Generative Algorithms}\label{sec:benchmark}
We put forward a framework for comparing \textit{tensorial representations} of 3D objects. To ensure a fair and meaningful comparison, the following assumptions and requirements are established:
\begin{itemize}
    \item \textbf{Meshes as ground truth:} The target representation in this study are \textit{meshes},  due to their fundamental role in 3D computer graphics. They facilitate rapid rendering and are highly efficient in terms of space. Prominent datasets, such as ShapeNet~\citep{chang2015shapenet} and Objaverse~\citep{deitke2023objaverse}, also provide objects as meshes.
    \item \textbf{Modularity:} The pipeline should be \textit{modular}, allowing for the integration of different representations while controlling other components.
    \item \textbf{Coverage of related work:} The framework is designed to align with state-of-the-art methods in the field, accommodating the most significant representations utilized in recent research.
\end{itemize}

Based on these features, we propose a unified pipeline that captures the essential elements of contemporary 3D generation methods. 
As depicted in \autoref{fig:pipeline}, our multi-stage generation pipeline involves the following stages:
\begin{enumerate} 
   \item \textit{Representation Conversion}: Transforming a 3D mesh into a suitable representation such as a voxel grid, SDF, or point cloud. This step often employs algorithmic transformations without requiring training.
   \item \textit{Representation Compression}: Utilizing an autoencoder to compress the high-dimensional representation into a lower-dimensional latent space using architectures like Autoencoder (AE)~\citep{kim2023neuralfield}, a Variational Autoencoder (VAE)~\citep{luo2021diffusion, ren2024xcube}, or a Vector Quantized Variational Autoencoder (VQ-VAE)~\citep{cheng2023sdfusion}.
   \item \textit{Latent Generation}: Training a generative model, typically a diffusion model~\citep{sohl2015deep, ho2020denoising, lyu2021conditional, zhou20213d}, to produce latent vectors that the decoder can reconstruct. If the representation is tokenizable, autoregressive models~\citep{yan2022shapeformer, mittal2022autosdf, siddiqui2023meshgpt} may be employed instead.
   \item \textit{Mesh Reconstruction}: Reconstructing the final 3D mesh from the decoded representation, often using algorithms like Marching Cubes~\citep{lorensen1987marching}.
\end{enumerate}

\begin{table}[h!]
\centering
\caption{Overview of representations, autoencoders and generators used in influential contemporary research on 3D generation (PC: Point cloud). Optimization-based methods are not included since they follow a different structure. Some methods do not compress the representation and directly denoise on the representation.
}
\label{tab:literature}
\resizebox{0.9\linewidth}{!}{
\begin{tabular}{lrrr}
\toprule
Paper & Representation & Compression & Generation \\
\midrule
gDNA~\citep{chen2022gdna} & Occupancy Field & - & GAN \\
BlockGAN~\citep{nguyen2020blockgan} & Voxel Grid & - & GAN  \\
XCube~\citep{ren2024xcube} & Voxel Grid & VAE & Sparse UNet diff.  \\
Neuralfield-LDM~\citep{kim2023neuralfield} & Voxel & AE (CNN) & Hierarchical diff. \\ 
Trellis~\citep{xiang2024structured} & Voxel / SparseFlex & Transformer VAE & Rectified Flow \\
\midrule
LAS-Diffusion~\citep{zheng2023locally} & Voxel / SDF & - & Diffusion UNet  \\ 
One-2-3-45++~\citep{liu2023onepp} & Voxel / SDF & - & Diffusion UNet \\ 
Diffusion-SDF~\citep{li2023diffusion} & Voxelized SDF & Patch-VAE & Diffusion UNet \\
LDM~\citep{xie2024ldm} & Voxelized SDF & - & Transformer  \\
SDFusion~\citep{cheng2023sdfusion} & Voxelized SDF & VQ-VAE & Diffusion UNet  \\
DualOctreeGNN~\citep{wang2022doctree} & Octree / SDF & AE \\
Make-A-Shape~\citep{hui2024make} & SDF & Wavelet features & Diffusion ViT\\
\midrule
Cannonical mapping~\citep{cheng2022autoregressive} & PC & VAE & Autoregressive  \\
DPM~\citep{luo2021diffusion} & PC & VAE & Diffusion  \\
PVD~\citep{zhou20213d} & Voxel / PC & - & Diffusion CNN 
\\
r-GAN / l-GAN~\citep{achlioptas2018learning} & PC & AE & GAN \\
SoftFlow~\citep{kim2020softflow} & PC & AE & Normalizing Flow \\
DPF-Net~\citep{klokov2020discrete} & PC & AE & Normalizing Flow \\
Shape-GF~\citep{cai2020learning} & PC / Density Field & AE & GAN \\
PointFlow~\citep{yang2019pointflow} & PC & VAE & Normalizing Flow \\
PointGrow~\citep{sun2020pointgrow} & PC & MLP & Autoregressive \\
3dAAE~\citep{zamorski2020adversarial} & PC & VAE & AAE \\
tree-GAN~\citep{shu20193d} & PC & -  & GAN \\
Point-E~\citep{nichol2022point} & PC & Coordinates & DiT  \\
3DShape2VecSet~\citep{zhang20233dshape2vecset} & PC / SDF & Transformer  & DiT \\
CLAY~\citep{zhang2024clay} & PC / SDF & Transformer
 & DiT \\
 Michelangelo~\citep{zhao2023michelangelo} & PC / Occupancy & VAE & Diffusion UNet \\
 SparseFlex~\citep{he2025sparseflex} & PC / SparseFlex & Transformer VAE & Rectified Flow \\ 
 \midrule
AutoSDF~\citep{mittal2022autosdf} & SDF & VQ-VAE & Autoregressive  \\ 
SurfGen~\citep{luo2021surfgen} & SDF & - & GAN \\
3D-LDM~\citep{nam20223d} & SDF & MLP & Diffusion (MLP) \\
TripoSG~\citep{li2025triposg} & SDF & Transformer VAE & Rectified Flow \\
\midrule
SDM-NET~\citep{gao2019sdm} & Mesh & - & VAE \\
PolyGen~\citep{nash2020polygen} & Mesh & Coordinates & Autoregressive \\
MeshGPT~\citep{siddiqui2023meshgpt} & Mesh (face tokens) & GraphAE & Autoregressive \\
MeshXL~\citep{chen2024meshxl} & Mesh & Coordinates & Autoregressive \\
\midrule
GIRAFFE~\citep{fu2022representing} & NeRF & - & GAN \\
HoloDiffusion~\citep{karnewar2023holodiffusion} & NeRF & ResNet & Diffusion  \\
Shape-E~\citep{jun2023shap} & NeRF / SDF & Transformer & DiT  \\
SSDNeRF~\citep{chen2023single} & NeRF / Triplane & MLP & DiT \\
\midrule
EG3D~\citep{chan2022efficient} & Triplane & - & GAN \\
3DGen~\citep{gupta20233dgen} & Triplane & PointNet/Unet & Diffusion UNet  \\
Direct3D~\citep{wu2024direct3d} & PC / Triplane & -  & DiT  \\
Get3D~\citep{gao2022get3d} & Triplane+DMTet & - & StyleGAN  \\
TriFlow~\citep{wizadwongsa2024taming} & Triplane & MLP & DiT \\
ShapeFormer~\citep{yan2022shapeformer} & PC / VQDIF & Transformer & Autoregressive \\
\bottomrule
\end{tabular}
}
\vspace{-5mm}
\end{table}

To support the claim that this pipeline is representative for most of the influential \textit{direct} 3D generation approaches published over the past years, we list related work and their instantiations of representation, compressor and generator in \autoref{tab:literature}. For instance, SDFusion~\citep{cheng2023sdfusion} compresses an SDF-grid with a VQ-VAE and denoises the latent with a U-Net-based diffusion model; XCube~\citep{ren2024xcube} encodes a voxel grid with a VAE and uses a multi-resolution U-Net diffusion; and Point-E~\citep{nichol2022point} compresses a point cloud and denoises it via a Diffusion Transformer (DiT). For an in-depth discussion of related work, we refer to \aref{sec:representations}.

\subsection{Diffusion models}

A suitable tensorial representation of 3D objects is part of any deep learning generative approach, whether it employs GANs, diffusion or autoregressive generation. Given their widespread use in the field, we opt for diffusion in our experiments; however, the main findings and identified shortcomings of certain representations are generic. We implement two diffusion models: a Diffusion Transformer (DiT)~\citep{peebles2023scalable} and a U-Net diffusion model~\citep{rombach2022high, cheng2023sdfusion}. DiTs are adept at handling high-dimensional data by utilizing self-attention mechanisms and can be applied to any representation with an appropriate tokenization scheme, making them suitable for representations like point clouds and meshes. Conversely, U-Net architectures excel with grid-based representations such as voxel grids and SDFs, where the data can be organized into 2D or 3D tensors. For a detailed technical introduction to diffusion processes and for implementation details, see \aref{app:diffusion}.

\subsection{Selected representations and architectures}\label{sec:selected_representations}

The experiments in this study are conducted using a carefully selected set of 3D representations, chosen for several key reasons. First, we prioritized representations that are widely used in the field, ensuring the relevance of our analysis and alignment with existing literature. 
Second, we aimed to include a diverse range of representations, such as point clouds used as intermediate representation in the training process, and implicit representations like density fields and SDFs, which we use as the output representation to facilitate conversion to meshes for all methods.
Third, scalability to large datasets and compatibility with diffusion models was a critical consideration, as this ensures that each representation can be effectively integrated with modern generative techniques. 
The encoders and decoders for each representation are trained through a reconstruction task on the same dataset used for the generative task. 
To ensure normally distributed latents, which is a crucial prerequisite for the diffusion process, the AE encoder always implements LayerNorm as its final layer. 
We implemented our representations, encoders, and diffusion models within a standardized training pipeline utilizing the \texttt{hydra} and \texttt{accelerate} libraries. 
\paragraph*{Voxel and SDF Grid Encoding}

Voxel grids offer an intuitive and explicit encoding for 3D objects. While used predominantly in early research~\citep{wu20153d, maturana2015voxnet, choy20163d, wu2016learning, brock2016generative, dai2018scancomplete}, they are still used in SOTA work~\citep{ren2024xcube, zheng2023locally, liu2023onepp}. Instead of binary occupancy indicators, some works fill the grid cells with sampled SDF values~\citep{cheng2023sdfusion, mittal2022autosdf} to increase the expressiveness. We implement both approaches: a standard voxel grid where each cell holds a binary occupancy value and an SDF grid where each cell contains the signed distance to the nearest point on the mesh surface, sampled from a truncated SDF with a cutoff of $0.2$. Meshes were converted to 3D grids of resolution $64^3$ by ray casting and closest point queries. For encoding, we implemented a 3D CNN following \citet{cheng2023sdfusion}. For example, a grid of resolution $64^3$ is transformed into a latent tensor of shape $3 \times 16 \times 16 \times 16$. This latent representation is typically diffused using a 3D U-Net~\citep{cheng2023sdfusion}. Here, we additionally introduce a transformer-based approach, where the latent tensor is tokenized by dividing it into 3D patches of size $4^3$, analogous to patch-based tokenization used in vision transformers. This results in a total of 12 potential generative approaches: Voxel / SDF grid, each combined with AE / VAE / VQ-VAE, and denoised with DiT or U-Net. 

\paragraph*{3DShape2VecSet Encoding}

Occupancy Networks~\citep{mescheder2019occupancy} and their extensions~\citep{peng2020convolutional, atzmon2020sal, zhang20233dshape2vecset} use implicit representations to sample density or occupancy at any spatial point, enabling detailed surface generation beyond grid limitations. This method has recently been leveraged for conditional 3D generation, achieving remarkable results~\citep{zhang2024clay, yang2024hunyuan3d, li2025triposg}. 
Our implementation, based on the 3DShape2VecSet model~\citep{zhang20233dshape2vecset}, transforms meshes into point clouds  $\boldsymbol{X}$ and sub-samples them ($\boldsymbol{\hat{X}}$). The positional embeddings of the points are processed through cross-attention between $\boldsymbol{X}$ and $\boldsymbol{\hat{X}}$, resulting in a set of $k$-dimensional latent vectors. The decoder passes the vector set through multiple self-attention layers, with cross-attention applied between the outputs and embedded query points. Following \citet{zhang20233dshape2vecset}, latent generation employs a DiT.

\paragraph*{Dual Octree Graph Encoding}

To efficiently capture high-resolution geometric details while maintaining computational scalability, we implemented the dual octree graph representation proposed by \citet{wang2022dual}. This method leverages the hierarchical structure of octrees combined with graph neural networks (GNNs) to effectively represent complex 3D geometries.
The dual octree graph network takes a set of point clouds as input, builds a dual octree graph, and outputs an adaptive feature volume via a graph-CNN-based encoder-decoder network structure. In all experiments, we use a tree depth of 6 and force the octree to be full when depth $d \leq 2$. 
During decoding, the encoded Octree-features are converted to a signed distance field via a multilevel neural partition-of-unity (Neural MPU) module to construct a surface. Our approach extended this AE structure to VAE and VQ-VAE and trained a simple two-layer 3D U-Net diffusion model for generation. To the best of our knowledge, this is the first generative approach based on Dual Octrees.

\paragraph*{Triplanes}

Triplane representations efficiently encode 3D scenes in a \textit{hybrid explicit-implicit} manner~\citep{wangdiffusion}, utilizing three orthogonal 2D feature planes~\citep{peng2020convolutional}. From these planes, a lightweight MLP occupancy network can perform volumetric rendering. 
We employ the autoencoder architecture proposed in~\citep{peng2020convolutional} to generate triplane latents from point clouds sampled on the mesh. Following ~\citet{wu2024blockfusion}, latent generation employs a UNet with 3D-aware convolution layers to better capture the unique properties of the triplane latents.

\paragraph*{NeRFs}

Neural radiance field (NeRF) is another implicit representation that was first proposed for novel view synthesis \citep{mildenhall2021nerf}, parameterizing the color and density at each point by a neural network.
To render the 3D object from novel views, a differentiable volumetric ray casting approach is utilized. The vast majority of generative methods using a NeRF representation are optimization-based, iteratively updating the parameters of the neural network to maximize the likelihood of rendered images under an image generation model \citep{poole2022dreamfusion, lin2023magic3d, liu2023zero, wang2024prolificdreamer, di2024boost}. 
Therefore, in this work, we adopt instead the NeRF representation used in Shap-E \citep{jun2023shap}, which falls into the framework illustrated in \autoref{fig:pipeline}. Each mesh is converted into a point cloud and multi-view RGB images, which are encoded into a latent vector using point convolution, cross attention, and a Transformer. To ensure that the latent vector is bounded, it is input into a tanh activation function, and noise is added; however, to be consistent with the other representations, we replace this step with layer normalization. During decoding, the latent vector is projected to the parameters of the NeRF's neural network. The autoencoder is trained on a combination of RGB and transmittance reconstruction losses on multi-view renderings. In the original paper, the latent vector is also projected to the parameters of a signed texture field, and the autoencoder is fine-tuned with SDF and color distillation losses. However, we exclude this step to have a fair comparison of NeRF with the other representations.
As in the original paper, we leverage DiT for generation. For both reconstruction and generation, meshes are obtained from the NeRF by applying Marching Cubes to its density function.

\subsection{Evaluation protocol: Joint evaluation of reconstruction and generation}\label{sec:methods_evaluation}

The ability to transform a mesh to and from latent space upper-bounds a representation's generation capabilities, as generated vectors must ultimately be converted into meshes. Thus, we advocate for jointly benchmarking reconstruction and generation to understand how reconstruction errors limit generation. Reconstruction error comprises representation-inherent errors, like information loss in voxel grids and compression errors due to the encoder. Additionally, an encoder trained on limited data may struggle with novel objects, necessitating an assessment of out-of-distribution (OOD) performance. Consequently, we propose the following evaluation protocol:
\begin{enumerate} 
    \item Assess reconstruction (mesh $\rightarrow$  representation $\rightarrow$ mesh) 
    \item Benchmark compression performance (mesh $\rightarrow$ representation $\rightarrow$ latent vector $\rightarrow$ representation $\rightarrow$ mesh). 
    \item Test the encoder's generalization in OOD tasks
    \item Measure generation performance quantitatively and    qualitatively.
\end{enumerate}

\section{Experimental setup}\label{sec:experimetnal_setup}

We benchmarked the selected methods on a subset of categories of the ShapeNet dataset \citep{chang2015shapenet}, specifically \emph{car}, \emph{airplane}, and \emph{chair}, following \citet{ren2024xcube}. The dataset was split into training, validation, and test sets according to the official division\footnote{\url{http://shapenet.cs.stanford.edu/shapenet/obj-zip/SHREC16/all.csv}}.  
We evaluated reconstruction performance on the full test dataset (806 objects for \emph{airplane}, 703 for \emph{car}, and 1,163 for \emph{chair}).
Following \citet{peng2020convolutional}, we assessed reconstruction quality using Chamfer Distance (CD), F-Score, and Normal Consistency (NC). For precise definitions, please refer to \aref{app:metrics}. The selected metrics can be directly computed on the meshes and, unlike metrics like Intersection over Union (IoU), do not assume meshes being manifold, watertight, or convertible to grid representations --- assumptions that do not hold for many meshes in the dataset.

\begin{figure*}[tbh]
    \centering
    \adjustbox{width=\textwidth}{
        \includegraphics[height=3cm]{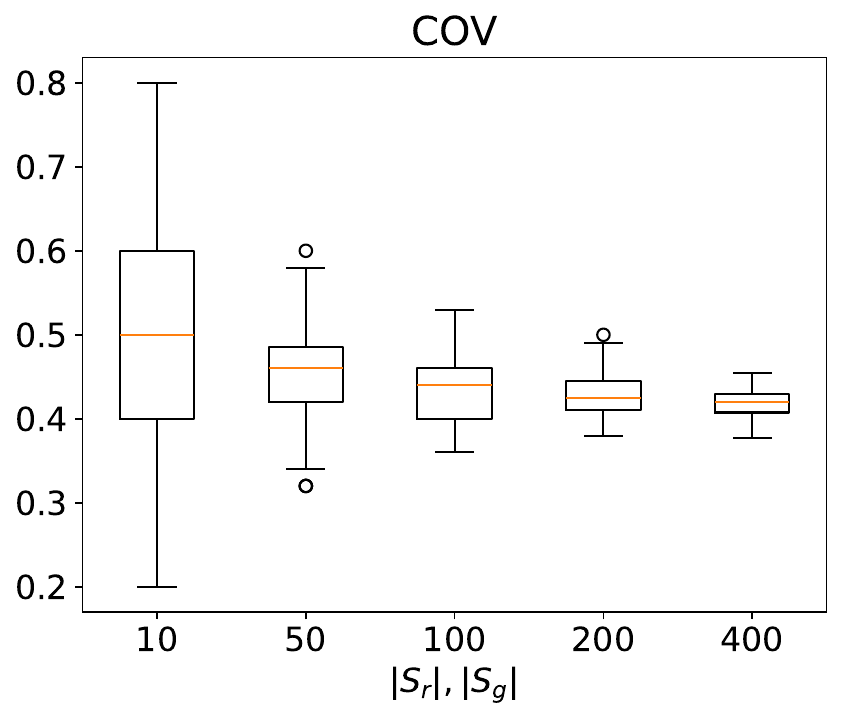}
        \includegraphics[height=3cm]{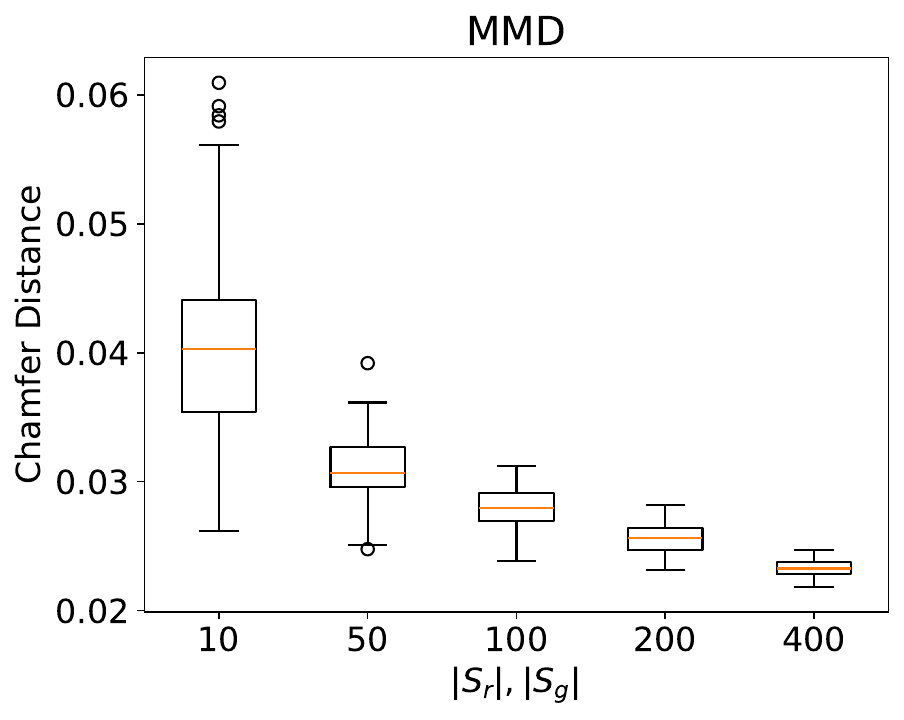}
        \includegraphics[height=3cm]{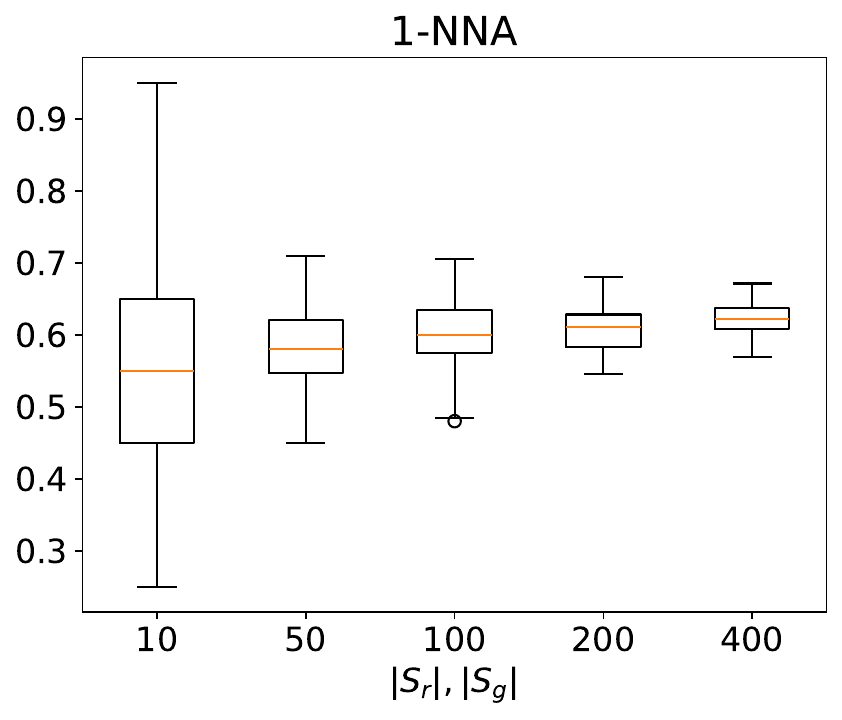}
    }
    \vspace{-5mm}
    \caption{We select 100 random subsets of different sizes of the train and test split of the ShapeNet airplane category and compute the COV, MMD, and 1-NNA metrics. 
    Small set sizes result in a large spread of values. We further observe that the Coverage (COV) is not close to the optimal value $1.0$ and the 1-Nearest Neighbor Accuracy (1-NNA) is above the optimal value $0.5$, indicating that learning from the training set may lead to biases.
    }
    \label{fig:shapenet_gt_uncond_metrics}
\end{figure*}

Quantitatively evaluating generative models for 3D shapes is challenging due to the diversity of tasks and metrics in related work, such as single-image or multi-view reconstruction, text-to-3D synthesis, or unconditional generation. To ensure a fair comparison and to avoid adding complexity to our benchmarking pipeline, we opted to evaluate the models in an unconditional generation setting. This approach tests the ability of the representations to generate objects that are both similar to the training set and diverse.
Following previous work \citep{gao2022get3d, lei2023nap, yang2019pointflow}, we implemented three distributional metrics: Coverage (COV), Minimum Matching Distance (MMD), and 1-Nearest Neighbor Accuracy (1-NNA). All these metrics are based on the pairwise CD between a set of generated samples $S_g$ and a reference dataset $S_r$, which is a subset of the test data. Coverage measures the fraction of $S_r$ that is matched to $S_g$, reflecting the diversity of the generated samples (higher is better). MMD computes the average minimum distance from samples in $S_r$ to those in $S_g$, indicating the quality of the generated samples (lower is better). The 1-NNA metric assesses the overfitting of the model by measuring the accuracy of classifying samples into $S_r$ and $S_g$ based on their distances. An ideal model achieves a 1-NNA of $0.5$, indicating that generated samples are indistinguishable from real data. For formal definitions of these metrics, see \aref{app:metrics}. 
In \autoref{fig:shapenet_gt_uncond_metrics}, we measure the stability of these metrics by computing them on random subsets of ShapeNet, imitating a perfect generative model. 
The meaningfulness of the metrics strongly depends on the chosen set size for $S_r$ and $S_g$, making it difficult to compare numbers in the literature. For instance, the MMD metric decreases with larger set sizes as the pool of meshes to find the closest neighbor increases.
We also observe a large spread of values for small set sizes and recommend that the set size should be larger than 200 for the metrics to converge.

To evaluate the subjective mesh quality, we conduct a user study. In each question, two meshes generated by different approaches are presented to the user, who is asked to indicate which is preferable based on object complexity and surface quality. Overall, we collect a dataset of 575 preferences from 24 users. To obtain scores for each approach, we model the preferences using the Bradley-Terry statistical model \citep{bradleyterry} (see \aref{app:user_study} for details).

\section{Results and Discussion}\label{sec:results}

\begin{figure}
    \centering
    \begin{minipage}{0.6\linewidth}
        \includegraphics[width=\linewidth]{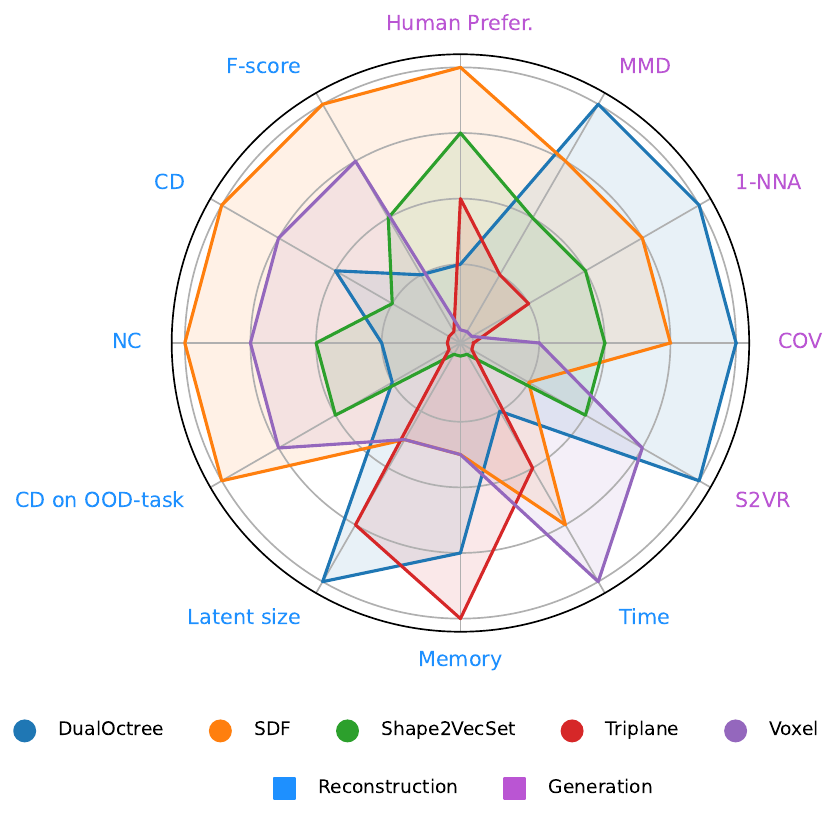}
    \end{minipage}%
    \hfill
    \begin{minipage}{0.3\linewidth}
        \caption{%
            Rankings of 3D representations based on generation and reconstruction metrics. 
            The outer circle indicates the top rank. This chart compares reconstruction metrics 
            (CD, F-score, NC), reconstruction generalization on an OOD task, and reconstruction 
            efficiency (memory footprint, encoding size, and inference time) as well as the 
            metrics for unconditional generation performance with user study rankings (1-NNA, 
            MMD, COV, surface-to-volume ratio (S2VR), Human Preference). 
        }
        \label{fig:spidergraph}
    \end{minipage}
\end{figure}

\autoref{fig:spidergraph} summarizes the results, demonstrating that the SDF-encoder achieves the best reconstruction performance, but our DualOctree diffusion model excels in generation. In the following, the results will be discussed step-by-step.

\subsection{Benchmarking generation performance}

The comparison of the best generative approaches, one per representation, is shown in \autoref{tab:generation}. For ablation studies on encoders, i.e. selecting the best from AE / VAE / VQVAE, and on diffusion models (DiT / UNet), see \aref{app:ablation_generation}. Interestingly, we found approaches with AE encoders usually outperforming VAEs, contrary to prior findings. Apparently, adding layer normalization to AE has a similar positive effect as the KL-loss in VAE. 
Surprisingly, the novel DualOctree-based diffusion model achieves best performance in all metrics, followed by SDF and Shape2VecSet. 
This finding contrasts with the widespread use of Shape2VecSet in state-of-the-art (SOTA) research, highlighting the critical role of multi-scale input schemes and the larger transformer models employed in recent SOTA developments~\citep{zhang2024clay}.

\begin{table}[ht]
\centering
\caption{Performance of representations in unconditional generation setting.}
\label{tab:generation}
\resizebox{0.5\linewidth}{!}{
\begin{tabular}{lrrr}
\toprule
Method & COV $\uparrow$ & MMD $\downarrow$ & 1-NNA $\rightarrow$ 0.5 \\
\midrule
DualOctree VAE UNet & \textbf{0.365} & \textbf{0.031} & \textbf{0.824} \\
SDF AE DiT & 0.357 & 0.032 & 0.860 \\
Shape2VecSet & 0.344 & 0.033 & 0.864 \\
Triplane AE UNet & 0.297 & 0.036 & 0.921 \\
Voxel AE DiT & 0.319 & 0.040 & 0.937 \\
\bottomrule
\end{tabular}
}
\end{table}

\begin{figure}
    \centering
    \includegraphics[width=0.9\linewidth]{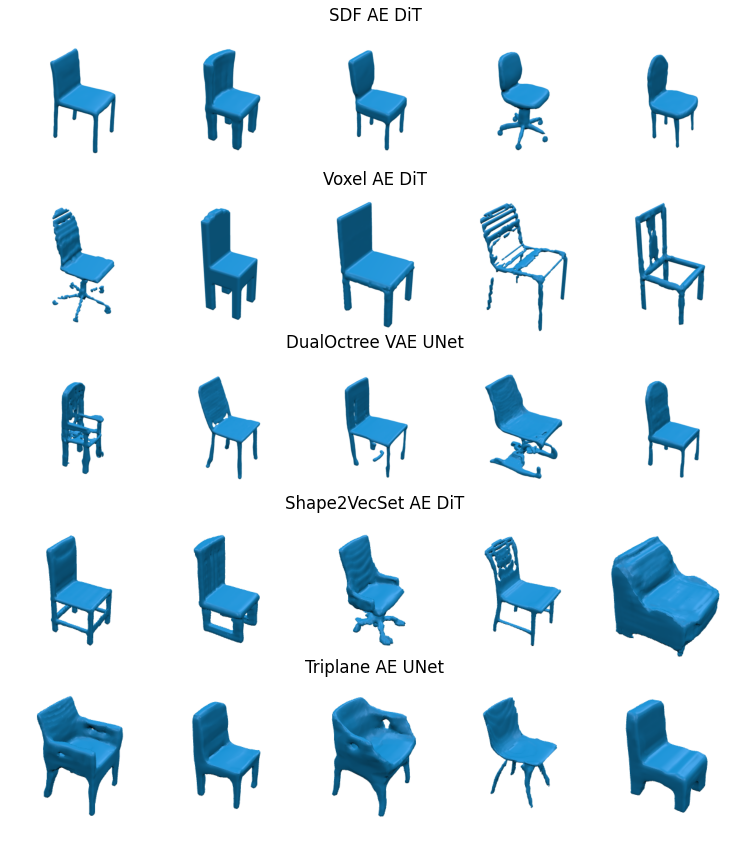}
    \caption{Qualitative results for mesh generation. We show results for each representation using the best encoder configuration for each.}
    \label{fig:generation_qualitative}
\end{figure}

However, the user study (see \autoref{fig:spidergraph}) paints a different picture, ranking SDF (score 0.25), Shape2vecset (-0.122) and Triplane (-0.140) above DualOctree (-0.178). This shows the necessity to run a user study when aiming to assess human-perceived object quality. One reason for the users' preference for SDF-generated objects may be the tendency of SDF-grids to generate smooth surfaces. \autoref{fig:generation_qualitative} provides qualitative results for the generative approaches. On the other hand, we found that DualOctree generates more \textit{complex} -- but potentially imperfect -- assets, measured in terms of the surface-to-volume ratio. \aref{app:complexity} expands on the analysis of complexity and provides further evidence that SDF tends to smoothen surfaces whereas Shape2VecSet and DualOctree are more prone to creating artifacts. Furthermore, \aref{app:objaverse} provides the corresponding results on a subset of the Objaverse dataset, demonstrating similar results but better performance of the Triplane representation.

\subsection{Reconstruction performance}

\autoref{tab:reconstruction} provides the reconstruction quality in terms of CD, F-Score, and NC. The best-vectorized representations are the voxel and SDF grid encodings using AE and VAE, with an average F-score of $88\%$. NeRF-encoding performs worst, probably due to modifications in our implementation for the sake of comparability (see section 2.2), and NeRF-based generation did not reach comparable generation performance and we thus only analyze its reconstruction performance.  
Ablation studies with respect to the encoder model (AE / VAE / VQ-VAE) are provided in \aref{app:ablation_recon}.

\begin{table}[h]
\caption{Reconstruction quality in terms of CD, F-score and NC. 
\label{tab:reconstruction}
}
\resizebox{\textwidth}{!}{
\begin{tabular}{l|lll|lll}
\toprule
 & \multicolumn{3}{c}{Trained \& tested on Airplane / Car / Chair}  & \multicolumn{3}{c}{OOD (trained on Chair, tested on Airplane)}  \\
\midrule
Method  & F-score (0.0125) $\uparrow$ & CD (*1e-4) $\downarrow$ & NC $\uparrow$ & F-score (0.0125) $\uparrow$ & CD (*1e-4) $\downarrow$ & NC $\uparrow$ \\
\midrule
DualOctree VAE & 76.122 \footnotesize{ $\pm$ 13.44} & 0.02 \footnotesize{ $\pm$ 0.01} & 0.766 \footnotesize{ $\pm$ 0.07} & 48.38 \footnotesize{ $\pm$ 11.39} & 0.047 \footnotesize{ $\pm$ 0.02} & 0.677 \footnotesize{ $\pm$ 0.08}  \\
NeRF AE & 58.44 \footnotesize{ $\pm$ 13.22} & 0.034 \footnotesize{ $\pm$ 0.02} & 0.723 \footnotesize{ $\pm$ 0.07} & 26.229 \footnotesize{ $\pm$ 11.64} & 0.107 \footnotesize{ $\pm$ 0.04} & 0.589 \footnotesize{ $\pm$ 0.05}  \\
SDF AE & \textbf{88.434 \footnotesize{ $\pm$ 6.58}} & \textbf{0.012 \footnotesize{ $\pm$ 0.0}} & \textbf{0.827 \footnotesize{ $\pm$ 0.06}} & \textbf{91.123 \footnotesize{ $\pm$ 6.02}} & \textbf{0.01 \footnotesize{ $\pm$ 0.01} }& \textbf{0.843 \footnotesize{ $\pm$ 0.05}} \\
Shape2VecSet AE & 79.37 \footnotesize{ $\pm$ 17.04} & 0.023 \footnotesize{ $\pm$ 0.02} & 0.776 \footnotesize{ $\pm$ 0.07} & 75.338 \footnotesize{ $\pm$ 8.87} & 0.022 \footnotesize{ $\pm$ 0.01} & 0.717 \footnotesize{ $\pm$ 0.07} \\
Triplane AE & 66.445 \footnotesize{ $\pm$ 16.06} & 0.028 \footnotesize{ $\pm$ 0.02} & 0.759 \footnotesize{ $\pm$ 0.08} & 41.69 \footnotesize{ $\pm$ 11.57} & 0.073 \footnotesize{ $\pm$ 0.03} & 0.688 \footnotesize{ $\pm$ 0.07}  \\
Voxel AE & 85.666 \footnotesize{ $\pm$ 10.54} & 0.016 \footnotesize{ $\pm$ 0.01} & 0.787 \footnotesize{ $\pm$ 0.06} & 85.602 \footnotesize{ $\pm$ 9.48} & 0.017 \footnotesize{ $\pm$ 0.01} & 0.8 \footnotesize{ $\pm$ 0.05} \\
\bottomrule
\end{tabular}
}
\vspace{-3mm}
\end{table}

Furthermore, we hypothesized that larger latent representations reduce the information loss during compression but come along with longer runtimes. \autoref{fig:reconstruction_memory} shows the trade-off between reconstruction loss, reconstruction runtime, and the size of the latent vector (in terms of the number of elements of the tensor). 
The results shows well the different characteristics and importance of different representations. For instance, DualOctree with the smallest latent size achieves better reconstructions than Triplane with a larger latent thanks to the spatial adaptive structure but is less accurate and slower than the Voxel representation that uses a much larger latent. Selecting the right representation can therefore significantly influence the runtime, accuracy and memory requirements. Generative methods targeting CAD applications may want a representation with minimal reconstruction error, while methods running on devices with small memory may favor a representation with a small latent.

\subsection{Generalization ability of the encoder}

When encoding 3D meshes into low-dimensional vectors, there is a trade-off between the compression capacity and the model's capability to generalize to new representations. Since the generator aims to design new objects, applicability beyond the training data is crucial. We quantified the generalization capability (see \autoref{tab:reconstruction} - right) where we test the encoders trained on the \textit{Chair} category when applied to the \textit{Airplane} category. While most encoders show very high generalization performance, on par with category-specific training, NeRFs and DualOctrees stand out as representations that struggle with OOD data. In the former, the MLP-based latent may be prone to overfitting, while for the latter, the low dimensionality of the latent could be problematic. 

\begin{table}[ht]
\begin{minipage}[t]{0.51\textwidth}
    \includegraphics[width=\textwidth]{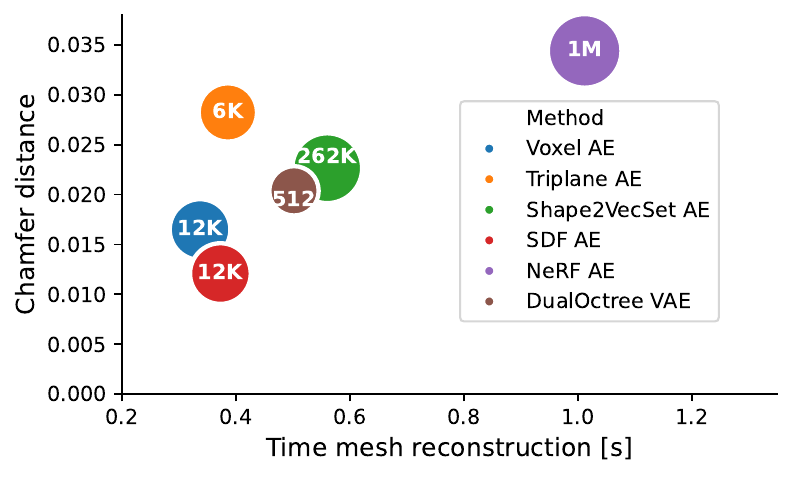}
    \captionof{figure}{Reconstruction quality by runtime for inference (decoding and mesh reconstruction), and  by size of the latent (see label and bubble size).}
    \label{fig:reconstruction_memory}
\end{minipage}
\hfill
\begin{minipage}[t]{0.47\textwidth}
    \includegraphics[width=\linewidth]{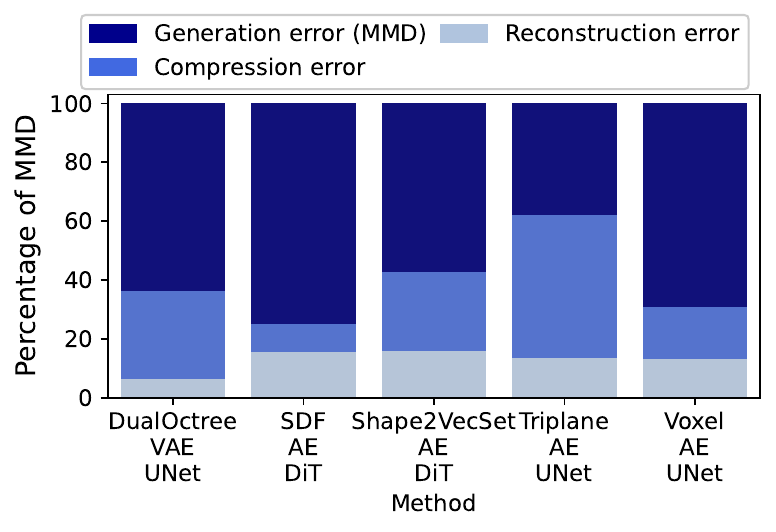}
    \captionof{figure}{Decomposition of errors during generation. Reconstruction errors and imperfect decompression play a substantial role.}
    \label{fig:gen_vs_recon_bars}

\end{minipage}
\vspace{-7mm}
\end{table}

\subsection{Effect of preprocessing}\label{sec:mesh_conversion}

\begin{figure}[htb]
    \centering
    \begin{subfigure}[b]{0.5\textwidth}
        \includegraphics[width=\linewidth]{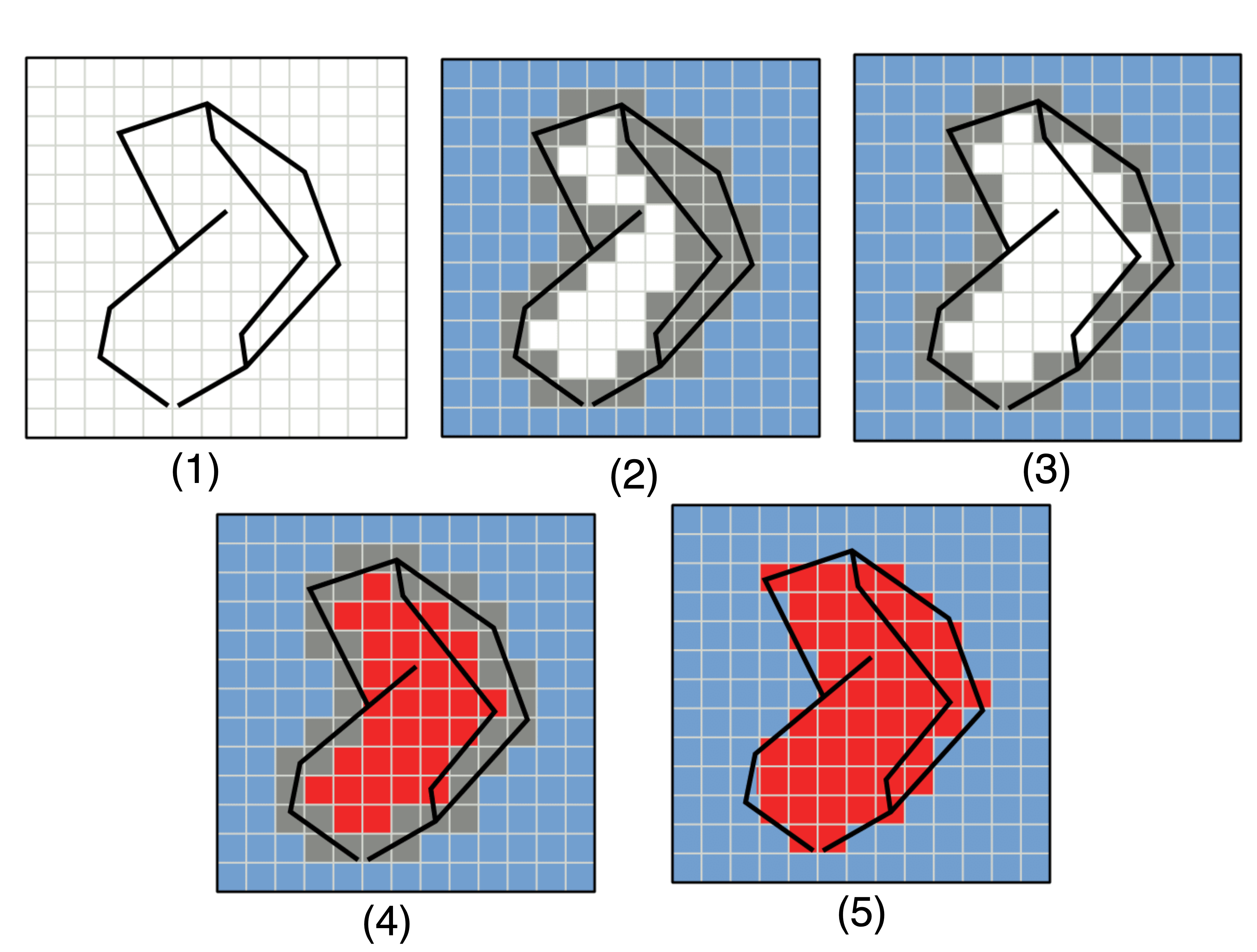}
        \caption{Chamfer Distance}
        \label{fig:flood_fill_alg}
    \end{subfigure}\hfill
    \begin{subfigure}[b]{0.47\textwidth}
        \includegraphics[width=\linewidth]{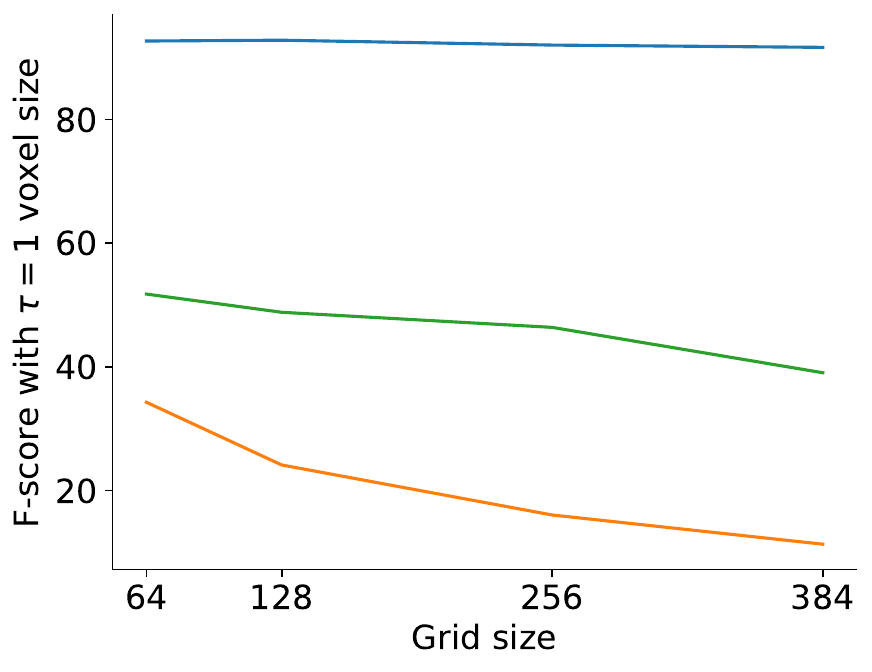}
        \caption{F-score}
        \label{fig:reconstruction_error_fscore}
    \end{subfigure}
    \caption{
    We compare different methods for converting a mesh to an SDF. \textbf{Ours} uses the flood fill approach to define the inside (red) and outside (blue) of a mesh, as explained in \subref{fig:flood_fill_alg}. 
    \textbf{Naive} uses a simple raycasting approach to determine the sign. \textbf{Mesh2SDF} from \citep{wang2022doctree} creates a watertight mesh using the unsigned distance and suffers from artificial thickening of the shape. This effect is reduced with increasing grid sizes.
    }
    \label{fig:reconstruction_error_mesh}
\end{figure}

For all methods — except those encoders operating directly on point clouds or meshes — preprocessing meshes is essential. Non-watertight meshes introduce significant artifacts as they do not have a well defined inside or outside, which is crucial information for many methods, adversely affecting the quality of the reconstructed shapes. Mesh manifoldization is an active research area, with various methods proposed like
the \textit{Manifold} library~\citep{huang2018robust}, which thickens surfaces to ensure that every mesh component forms a solid volume. While effective, this process can alter the original geometry by artificially expanding thin structures. \textit{ManifoldPlus}~\citep{huang2020manifoldplus} extends \textit{Manifold} but has been reported to produce inconsistent results \citep{zhang2024clay}. 
\citet{wang2022dual} uses a simple manifoldization (Mesh2SDF) based on contouring the unsigned distance function, which also produces significant thickening artifacts or requires high grid resolutions, resulting in high compute and memory requirements.
The thickening of the mesh helps to preserve thin structures but adds an irrevocable bias. To this end, we introduce a lean and mean preprocessing step that transforms meshes to SDFs without thickening. \autoref{fig:flood_fill_alg} shows a visual explanation of our preprocessing step using the flood fill algorithm. Instead of altering the mesh we define the inside and outside as outlined in \autoref{fig:flood_fill_alg}:
\begin{enumerate}
    \item We define a target grid encompassing the mesh. 
    \item We mark all voxels that touch the mesh surface (gray) and then flood fill starting from a corner voxel guaranteed to be outside to define the outside region (blue).
    \item Voxels that touch the mesh but not an outside voxel defined by the 26-voxel-neighborhood get removed. This step effectively eliminates internal structures not considered part of the outer shape of an object. \item All unlabelled voxels get labelled as \emph{inside} (red). 
    \item For all gray voxels we determine whether the voxel center lies inside or outside by comparing to the plane defined by the closest point to the surface and the normal approximated by the sum of the positions of the outside neighbor voxels. The surface points sampled in this step are reused for computing distances for the SDF. 
\end{enumerate}
Finally, we directly sample points and compute distances with the determined sign.

We compare the effects of manifoldization as a preprocessing step by converting meshes to a grid SDF representation and back to the mesh representation in \autoref{fig:reconstruction_error_fscore} (see \autoref{app:airplane_recon} for additional results). 

\subsection{Error decomposition}

In alignment with the goal to evaluate reconstruction and generation \textit{jointly}, 
we investigate the relation between generation, compression, and reconstruction errors. Since MMD measures generation performance as the mean CD between each sample in $S_r$ and its best match in $S_g$, it can be compared to the CD at the reconstruction or compression stage. While the generation of new shapes inevitably induces an MMD $> 0$ (see \autoref{fig:shapenet_gt_uncond_metrics}), the MMD is loosely lower bounded by the reconstruction and compression error in terms of CD. To decompose the error, for each sample $\rho \in S_r$, we compare its MMD, its compression error (e.g. encoding-decoding), and its mesh reconstruction error (e.g. marching cubes inaccuracies). The reconstruction errors are taken from the analysis in \autoref{fig:reconstruction_error_fscore} while the compression errors correspond to the sample-wise result from \autoref{tab:reconstruction}. It is worth noting that these errors are not simply additive; however, it is interesting to investigate the size of the lower bounds (reconstruction and compression errors) in relation to the absolute size of the generative error. Figure~\ref{fig:gen_vs_recon_bars} shows that the reconstruction and compression errors amount to 12.9\% and 39.3\% of the MMD, respectively, when averaging over all representations. The MMD and the reconstruction CD correlate on average with a Pearson R correlation of 0.30 ($p<0.01$). As expected, there is also a positive correlation of $r=0.24$ between compression and reconstruction error. This analysis underlines the important role of 3D representations and their reconstruction performance, as reconstruction errors can significantly reduce the quality of the generated samples.

\section{Conclusion}\label{sec:conclusion}
We have presented a systematic comparison of 3D representations for reconstruction and generation. Our analysis leads us to recommend the following best practices:

\begin{itemize}
    \item Evaluate reconstruction and generation jointly. Since generation quality is upper-bounded by the reconstruction error, the errors resulting from mesh reconstruction and representation encoding alone should be reported for transparency.
    \item Compute the errors with respect to the original mesh. Results should not be distorted by derivatives obtained following the application of methods like \textit{Manifold}\footnote{see \autoref{fig:gen_vs_recon_bars} and \autoref{fig:reconstruction_error_mesh} for the influence of preprocessing techniques}, and avoid metrics that require additional preprocessing.
    \item Include conversions of the output representation to meshes. Comparing only generated surface points to the ground truth mesh may skew results.
    \item A sufficient number of samples is crucial when evaluating unconditional generation with metrics such as MMD, 1-NNA, and Coverage. More than 200 samples are generally necessary to achieve robust outcomes.
\end{itemize}

\paragraph*{Limitations}
This framework covers \textit{direct} 3D generation, in contrast to optimization-based methods such as \textit{Score Distillation Sampling (SDS) }-based methods~\citep{poole2022dreamfusion, chen2023fantasia3d, lin2023magic3d, sun2023dreamcraft3d, wang2024prolificdreamer, shi2023mvdream} or other approaches that use multi-view images for inference~\citep{shi2023zero123++, liu2024one, long2024wonder3d}. Despite the impressive results generated with these techniques, recent developments have shifted the focus of the field back to direct 3D generation, due to 1) high computational costs of inference-time optimization~\citep{li2024advances}, 2) dependence of the generation quality on the fidelity of the multi-view images~\citep{wu2024direct3d}, 3) preferability of explicit 3D outputs for artists~\citep{yang2024hunyuan3d}, and 4) feasibility of training for direct generation due to large-scale 3D datasets such as Objaverse~\citep{deitke2023objaverse} and ObjaverseXL~\citep{deitke2024objaverse}. Similarly, Gaussian Splatting has become popular in the 3D field~\citep{tang2023dreamgaussian, kerbl20233d, lan2024gaussiananything}, but is not included in our study, as it is usually trained with SDS-loss methods and focuses on realistic and efficient rendering, in contrast to our assumption of meshes as the ground truth. 

Furthermore, some generative methods fit into our unified pipeline but remain unimplemented, such as autoregressive generation. Follow-up work could also extend the analysis to other datasets. While we already provide results on a subset of Objaverse in \aref{app:objaverse}, further experiments distinguishing different types of objects (e.g. thin structures or very complex shapes) could bring further insights on the advantages and drawbacks of each representation.

\paragraph{Outlook}

There are several open challenges in the field, such as accounting for interior structure of objects, transferring 3D \textit{object} generation methods to a \textit{scene} level~\citep{ren2024xcube}, or enabling model articulation~\citep{leboutet2024midgard, lei2023nap, liu2024cage}. While our benchmark shows the general ability of any representation to encode and generate high-resolution objects, it also shows the challenges such as dealing with thin structures while retaining computational efficiency. 
With the presented pipeline implemented in an open-source code base, we offer a unified framework to develop and benchmark novel 3D representations for generation.



\bibliography{references}

\begin{thebibliography}{108}
\providecommand{\natexlab}[1]{#1}
\providecommand{\url}[1]{\texttt{#1}}
\expandafter\ifx\csname urlstyle\endcsname\relax
  \providecommand{\doi}[1]{doi: #1}\else
  \providecommand{\doi}{doi: \begingroup \urlstyle{rm}\Url}\fi

\bibitem[Achlioptas et~al.(2018)Achlioptas, Diamanti, Mitliagkas, and Guibas]{achlioptas2018learning}
Panos Achlioptas, Olga Diamanti, Ioannis Mitliagkas, and Leonidas Guibas.
\newblock Learning representations and generative models for 3d point clouds.
\newblock In \emph{International conference on machine learning}, pages 40--49. PMLR, 2018.

\bibitem[Atzmon and Lipman(2020)]{atzmon2020sal}
Matan Atzmon and Yaron Lipman.
\newblock Sal: Sign agnostic learning of shapes from raw data.
\newblock In \emph{Proceedings of the IEEE/CVF conference on computer vision and pattern recognition}, pages 2565--2574, 2020.

\bibitem[Bradley and Terry(1952)]{bradleyterry}
Ralph~Allan Bradley and Milton~E Terry.
\newblock Rank analysis of incomplete block designs: I. the method of paired comparisons.
\newblock \emph{Biometrika}, 39\penalty0 (3/4):\penalty0 324--345, 1952.

\bibitem[Brock et~al.(2016)Brock, Lim, Ritchie, and Weston]{brock2016generative}
Andrew Brock, Theodore Lim, James~M Ritchie, and Nick Weston.
\newblock Generative and discriminative voxel modeling with convolutional neural networks.
\newblock \emph{arXiv preprint arXiv:1608.04236}, 2016.

\bibitem[Cai et~al.(2020)Cai, Yang, Averbuch-Elor, Hao, Belongie, Snavely, and Hariharan]{cai2020learning}
Ruojin Cai, Guandao Yang, Hadar Averbuch-Elor, Zekun Hao, Serge Belongie, Noah Snavely, and Bharath Hariharan.
\newblock Learning gradient fields for shape generation.
\newblock In \emph{Computer Vision--ECCV 2020: 16th European Conference, Glasgow, UK, August 23--28, 2020, Proceedings, Part III 16}, pages 364--381. Springer, 2020.

\bibitem[Cao et~al.(2024)Cao, Tan, Gao, Xu, Chen, Heng, and Li]{cao2024survey}
Hanqun Cao, Cheng Tan, Zhangyang Gao, Yilun Xu, Guangyong Chen, Pheng-Ann Heng, and Stan~Z Li.
\newblock A survey on generative diffusion models.
\newblock \emph{IEEE transactions on knowledge and data engineering}, 36\penalty0 (7):\penalty0 2814--2830, 2024.

\bibitem[Chan et~al.(2022)Chan, Lin, Chan, Nagano, Pan, De~Mello, Gallo, Guibas, Tremblay, Khamis, et~al.]{chan2022efficient}
Eric~R Chan, Connor~Z Lin, Matthew~A Chan, Koki Nagano, Boxiao Pan, Shalini De~Mello, Orazio Gallo, Leonidas~J Guibas, Jonathan Tremblay, Sameh Khamis, et~al.
\newblock Efficient geometry-aware 3d generative adversarial networks.
\newblock In \emph{Proceedings of the IEEE/CVF conference on computer vision and pattern recognition}, pages 16123--16133, 2022.

\bibitem[Chang et~al.(2015)Chang, Funkhouser, Guibas, Hanrahan, Huang, Li, Savarese, Savva, Song, Su, et~al.]{chang2015shapenet}
Angel~X Chang, Thomas Funkhouser, Leonidas Guibas, Pat Hanrahan, Qixing Huang, Zimo Li, Silvio Savarese, Manolis Savva, Shuran Song, Hao Su, et~al.
\newblock Shapenet: An information-rich 3d model repository.
\newblock \emph{arXiv preprint arXiv:1512.03012}, 2015.

\bibitem[Chen et~al.(2023{\natexlab{a}})Chen, Gu, Chen, Tian, Tu, Liu, and Su]{chen2023single}
Hansheng Chen, Jiatao Gu, Anpei Chen, Wei Tian, Zhuowen Tu, Lingjie Liu, and Hao Su.
\newblock Single-stage diffusion nerf: A unified approach to 3d generation and reconstruction.
\newblock In \emph{Proceedings of the IEEE/CVF international conference on computer vision}, pages 2416--2425, 2023{\natexlab{a}}.

\bibitem[Chen et~al.(2023{\natexlab{b}})Chen, Chen, Jiao, and Jia]{chen2023fantasia3d}
Rui Chen, Yongwei Chen, Ningxin Jiao, and Kui Jia.
\newblock Fantasia3d: Disentangling geometry and appearance for high-quality text-to-3d content creation.
\newblock In \emph{Proceedings of the IEEE/CVF international conference on computer vision}, pages 22246--22256, 2023{\natexlab{b}}.

\bibitem[Chen et~al.(2024{\natexlab{a}})Chen, Zhang, Liang, Luo, Li, Liu, Li, Long, Feng, and Tan]{chen2024dora}
Rui Chen, Jianfeng Zhang, Yixun Liang, Guan Luo, Weiyu Li, Jiarui Liu, Xiu Li, Xiaoxiao Long, Jiashi Feng, and Ping Tan.
\newblock Dora: Sampling and benchmarking for 3d shape variational auto-encoders.
\newblock \emph{arXiv preprint arXiv:2412.17808}, 2024{\natexlab{a}}.

\bibitem[Chen et~al.(2024{\natexlab{b}})Chen, Chen, Pang, Zeng, Cheng, Fu, Yin, Wang, Yu, Yu, Fu, and Chen]{chen2024meshxl}
Sijin Chen, Xin Chen, Anqi Pang, Xianfang Zeng, Wei Cheng, Yijun Fu, Fukun Yin, Billzb Wang, Jingyi Yu, Gang Yu, Bin Fu, and Tao Chen.
\newblock Meshxl: Neural coordinate field for generative 3d foundation models.
\newblock In \emph{NeurIPS}, 2024{\natexlab{b}}.

\bibitem[Chen et~al.(2022)Chen, Jiang, Song, Yang, Black, Geiger, and Hilliges]{chen2022gdna}
Xu~Chen, Tianjian Jiang, Jie Song, Jinlong Yang, Michael~J Black, Andreas Geiger, and Otmar Hilliges.
\newblock gdna: Towards generative detailed neural avatars.
\newblock In \emph{Proceedings of the IEEE/CVF Conference on Computer Vision and Pattern Recognition}, pages 20427--20437, 2022.

\bibitem[Cheng et~al.(2022)Cheng, Li, Liu, Sun, and Yang]{cheng2022autoregressive}
An-Chieh Cheng, Xueting Li, Sifei Liu, Min Sun, and Ming-Hsuan Yang.
\newblock Autoregressive 3d shape generation via canonical mapping.
\newblock In \emph{European Conference on Computer Vision}, pages 89--104. Springer, 2022.

\bibitem[Cheng et~al.(2023)Cheng, Lee, Tulyakov, Schwing, and Gui]{cheng2023sdfusion}
Yen-Chi Cheng, Hsin-Ying Lee, Sergey Tulyakov, Alexander~G Schwing, and Liang-Yan Gui.
\newblock {SDFusion}: Multimodal 3d shape completion, reconstruction, and generation.
\newblock In \emph{Proceedings of the IEEE/CVF Conference on Computer Vision and Pattern Recognition}, pages 4456--4465, 2023.

\bibitem[Choy et~al.(2016)Choy, Xu, Gwak, Chen, and Savarese]{choy20163d}
Christopher~B Choy, Danfei Xu, JunYoung Gwak, Kevin Chen, and Silvio Savarese.
\newblock 3d-r2n2: A unified approach for single and multi-view 3d object reconstruction.
\newblock In \emph{Computer Vision--ECCV 2016: 14th European Conference, Amsterdam, The Netherlands, October 11-14, 2016, Proceedings, Part VIII 14}, pages 628--644. Springer, 2016.

\bibitem[Crowson et~al.(2024)Crowson, Baumann, Birch, Abraham, Kaplan, and Shippole]{crowson2024scalable}
Katherine Crowson, Stefan~Andreas Baumann, Alex Birch, Tanishq~Mathew Abraham, Daniel~Z Kaplan, and Enrico Shippole.
\newblock Scalable high-resolution pixel-space image synthesis with hourglass diffusion transformers.
\newblock In \emph{Forty-first International Conference on Machine Learning}, 2024.

\bibitem[Dai et~al.(2018)Dai, Ritchie, Bokeloh, Reed, Sturm, and Nie{\ss}ner]{dai2018scancomplete}
Angela Dai, Daniel Ritchie, Martin Bokeloh, Scott Reed, J{\"u}rgen Sturm, and Matthias Nie{\ss}ner.
\newblock Scancomplete: Large-scale scene completion and semantic segmentation for 3d scans.
\newblock In \emph{Proceedings of the IEEE Conference on Computer Vision and Pattern Recognition}, pages 4578--4587, 2018.

\bibitem[Deitke et~al.(2023)Deitke, Schwenk, Salvador, Weihs, Michel, VanderBilt, Schmidt, Ehsani, Kembhavi, and Farhadi]{deitke2023objaverse}
Matt Deitke, Dustin Schwenk, Jordi Salvador, Luca Weihs, Oscar Michel, Eli VanderBilt, Ludwig Schmidt, Kiana Ehsani, Aniruddha Kembhavi, and Ali Farhadi.
\newblock Objaverse: A universe of annotated 3d objects.
\newblock In \emph{Proceedings of the IEEE/CVF Conference on Computer Vision and Pattern Recognition}, pages 13142--13153, 2023.

\bibitem[Deitke et~al.(2024)Deitke, Liu, Wallingford, Ngo, Michel, Kusupati, Fan, Laforte, Voleti, Gadre, et~al.]{deitke2024objaverse}
Matt Deitke, Ruoshi Liu, Matthew Wallingford, Huong Ngo, Oscar Michel, Aditya Kusupati, Alan Fan, Christian Laforte, Vikram Voleti, Samir~Yitzhak Gadre, et~al.
\newblock Objaverse-xl: A universe of 10m+ 3d objects.
\newblock \emph{Advances in Neural Information Processing Systems}, 36, 2024.

\bibitem[Di~Sario et~al.(2024)Di~Sario, Renzulli, Tartaglione, and Grangetto]{di2024boost}
Francesco Di~Sario, Riccardo Renzulli, Enzo Tartaglione, and Marco Grangetto.
\newblock Boost your nerf: A model-agnostic mixture of experts framework for high quality and efficient rendering.
\newblock In \emph{European Conference on Computer Vision}, pages 176--192. Springer, 2024.

\bibitem[Fridovich-Keil et~al.(2022)Fridovich-Keil, Yu, Tancik, Chen, Recht, and Kanazawa]{fridovich2022plenoxels}
Sara Fridovich-Keil, Alex Yu, Matthew Tancik, Qinhong Chen, Benjamin Recht, and Angjoo Kanazawa.
\newblock Plenoxels: Radiance fields without neural networks.
\newblock In \emph{Proceedings of the IEEE/CVF conference on computer vision and pattern recognition}, pages 5501--5510, 2022.

\bibitem[Fu et~al.(2022)Fu, Deng, Gao, and Zhang]{fu2022representing}
Tianlin Fu, Renyu Deng, Yue Gao, and Fuquan Zhang.
\newblock Representing scenes as compositional generative neural feature fields based on giraffe for 3d reconstruction of classroom scenes.
\newblock In \emph{International Conference on Intelligent Information Hiding and Multimedia Signal Processing}, pages 227--237. Springer, 2022.

\bibitem[Gao et~al.(2022)Gao, Shen, Wang, Chen, Yin, Li, Litany, Gojcic, and Fidler]{gao2022get3d}
Jun Gao, Tianchang Shen, Zian Wang, Wenzheng Chen, Kangxue Yin, Daiqing Li, Or~Litany, Zan Gojcic, and Sanja Fidler.
\newblock Get3d: A generative model of high quality 3d textured shapes learned from images.
\newblock \emph{Advances In Neural Information Processing Systems}, 35:\penalty0 31841--31854, 2022.

\bibitem[Gao et~al.(2019)Gao, Yang, Wu, Yuan, Fu, Lai, and Zhang]{gao2019sdm}
Lin Gao, Jie Yang, Tong Wu, Yu-Jie Yuan, Hongbo Fu, Yu-Kun Lai, and Hao Zhang.
\newblock Sdm-net: Deep generative network for structured deformable mesh.
\newblock \emph{ACM Transactions on Graphics (TOG)}, 38\penalty0 (6):\penalty0 1--15, 2019.

\bibitem[Gezawa et~al.(2020)Gezawa, Zhang, Wang, and Yunqi]{gezawa2020review}
Abubakar~Sulaiman Gezawa, Yan Zhang, Qicong Wang, and Lei Yunqi.
\newblock A review on deep learning approaches for 3d data representations in retrieval and classifications.
\newblock \emph{IEEE access}, 8:\penalty0 57566--57593, 2020.

\bibitem[Gupta et~al.(2023)Gupta, Xiong, Nie, Jones, and O{\u{g}}uz]{gupta20233dgen}
Anchit Gupta, Wenhan Xiong, Yixin Nie, Ian Jones, and Barlas O{\u{g}}uz.
\newblock 3dgen: Triplane latent diffusion for textured mesh generation.
\newblock \emph{arXiv preprint arXiv:2303.05371}, 2023.

\bibitem[He et~al.(2025)He, Zou, Chen, Guo, Liang, Yuan, Ouyang, Cao, and Li]{he2025sparseflex}
Xianglong He, Zi-Xin Zou, Chia-Hao Chen, Yuan-Chen Guo, Ding Liang, Chun Yuan, Wanli Ouyang, Yan-Pei Cao, and Yangguang Li.
\newblock Sparseflex: High-resolution and arbitrary-topology 3d shape modeling.
\newblock \emph{arXiv preprint arXiv:2503.21732}, 2025.

\bibitem[Hegde et~al.(2023)Hegde, Valanarasu, and Patel]{hegde2023clip}
Deepti Hegde, Jeya Maria~Jose Valanarasu, and Vishal Patel.
\newblock Clip goes 3d: Leveraging prompt tuning for language grounded 3d recognition.
\newblock In \emph{Proceedings of the IEEE/CVF International Conference on Computer Vision}, pages 2028--2038, 2023.

\bibitem[Ho et~al.(2020)Ho, Jain, and Abbeel]{ho2020denoising}
Jonathan Ho, Ajay Jain, and Pieter Abbeel.
\newblock Denoising diffusion probabilistic models.
\newblock \emph{Advances in neural information processing systems}, 33:\penalty0 6840--6851, 2020.

\bibitem[Huang et~al.(2018)Huang, Su, and Guibas]{huang2018robust}
Jingwei Huang, Hao Su, and Leonidas Guibas.
\newblock Robust watertight manifold surface generation method for shapenet models.
\newblock \emph{arXiv preprint arXiv:1802.01698}, 2018.

\bibitem[Huang et~al.(2020)Huang, Zhou, and Guibas]{huang2020manifoldplus}
Jingwei Huang, Yichao Zhou, and Leonidas Guibas.
\newblock Manifoldplus: A robust and scalable watertight manifold surface generation method for triangle soups.
\newblock \emph{arXiv preprint arXiv:2005.11621}, 2020.

\bibitem[Hui et~al.(2024)Hui, Sanghi, Rampini, Malekshan, Liu, Shayani, and Fu]{hui2024make}
Ka-Hei Hui, Aditya Sanghi, Arianna Rampini, Kamal~Rahimi Malekshan, Zhengzhe Liu, Hooman Shayani, and Chi-Wing Fu.
\newblock Make-a-shape: a ten-million-scale 3d shape model.
\newblock In \emph{Forty-first International Conference on Machine Learning}, 2024.

\bibitem[Jiang(2024)]{jiang2024survey}
Chenhan Jiang.
\newblock A survey on text-to-3d contents generation in the wild.
\newblock \emph{arXiv preprint arXiv:2405.09431}, 2024.

\bibitem[Jun and Nichol(2023)]{jun2023shap}
Heewoo Jun and Alex Nichol.
\newblock Shap-e: Generating conditional 3d implicit functions.
\newblock \emph{arXiv preprint arXiv:2305.02463}, 2023.

\bibitem[Karnewar et~al.(2023)Karnewar, Vedaldi, Novotny, and Mitra]{karnewar2023holodiffusion}
Animesh Karnewar, Andrea Vedaldi, David Novotny, and Niloy~J Mitra.
\newblock Holodiffusion: Training a 3d diffusion model using 2d images.
\newblock In \emph{Proceedings of the IEEE/CVF conference on computer vision and pattern recognition}, pages 18423--18433, 2023.

\bibitem[Kerbl et~al.(2023)Kerbl, Kopanas, Leimk{\"u}hler, and Drettakis]{kerbl20233d}
Bernhard Kerbl, Georgios Kopanas, Thomas Leimk{\"u}hler, and George Drettakis.
\newblock 3d gaussian splatting for real-time radiance field rendering.
\newblock \emph{ACM Trans. Graph.}, 42\penalty0 (4):\penalty0 139--1, 2023.

\bibitem[Kim et~al.(2020)Kim, Lee, Kang, Lee, and Kim]{kim2020softflow}
Hyeongju Kim, Hyeonseung Lee, Woo~Hyun Kang, Joun~Yeop Lee, and Nam~Soo Kim.
\newblock Softflow: Probabilistic framework for normalizing flow on manifolds.
\newblock \emph{Advances in Neural Information Processing Systems}, 33:\penalty0 16388--16397, 2020.

\bibitem[Kim et~al.(2023)Kim, Brown, Yin, Kreis, Schwarz, Li, Rombach, Torralba, and Fidler]{kim2023neuralfield}
Seung~Wook Kim, Bradley Brown, Kangxue Yin, Karsten Kreis, Katja Schwarz, Daiqing Li, Robin Rombach, Antonio Torralba, and Sanja Fidler.
\newblock Neuralfield-ldm: Scene generation with hierarchical latent diffusion models.
\newblock In \emph{Proceedings of the IEEE/CVF conference on computer vision and pattern recognition}, pages 8496--8506, 2023.

\bibitem[Klokov et~al.(2020)Klokov, Boyer, and Verbeek]{klokov2020discrete}
Roman Klokov, Edmond Boyer, and Jakob Verbeek.
\newblock Discrete point flow networks for efficient point cloud generation.
\newblock In \emph{European Conference on Computer Vision}, pages 694--710. Springer, 2020.

\bibitem[Lan et~al.(2024)Lan, Zhou, Lyu, Hong, Yang, Dai, Pan, and Loy]{lan2024gaussiananything}
Yushi Lan, Shangchen Zhou, Zhaoyang Lyu, Fangzhou Hong, Shuai Yang, Bo~Dai, Xingang Pan, and Chen~Change Loy.
\newblock Gaussiananything: Interactive point cloud flow matching for 3d object generation.
\newblock \emph{arXiv preprint arXiv:2411.08033}, 2024.

\bibitem[Leboutet et~al.(2024)Leboutet, Wiedemann, Cai, Paulitsch, and Yuan]{leboutet2024midgard}
Quentin Leboutet, Nina Wiedemann, Zhipeng Cai, Michael Paulitsch, and Kai Yuan.
\newblock Midgard: Modular interpretable diffusion over graphs for articulated designs.
\newblock In \emph{Advances in Neural Information Processing Systems}, 2024.

\bibitem[Lei et~al.(2023)Lei, Deng, Shen, Guibas, and Daniilidis]{lei2023nap}
Jiahui Lei, Congyue Deng, William~B Shen, Leonidas~J Guibas, and Kostas Daniilidis.
\newblock Nap: Neural 3d articulated object prior.
\newblock \emph{Advances in Neural Information Processing Systems}, 36:\penalty0 31878--31894, 2023.

\bibitem[Li et~al.(2023)Li, Duan, Zhou, and Lu]{li2023diffusion}
Muheng Li, Yueqi Duan, Jie Zhou, and Jiwen Lu.
\newblock Diffusion-sdf: Text-to-shape via voxelized diffusion.
\newblock In \emph{Proceedings of the IEEE/CVF conference on computer vision and pattern recognition}, pages 12642--12651, 2023.

\bibitem[Li et~al.(2024)Li, Zhang, Kang, Cheng, Gao, Zhang, Liang, Liao, Cao, and Shan]{li2024advances}
Xiaoyu Li, Qi~Zhang, Di~Kang, Weihao Cheng, Yiming Gao, Jingbo Zhang, Zhihao Liang, Jing Liao, Yan-Pei Cao, and Ying Shan.
\newblock Advances in 3d generation: A survey.
\newblock \emph{arXiv preprint arXiv:2401.17807}, 2024.

\bibitem[Li et~al.(2025)Li, Zou, Liu, Wang, Liang, Yu, Liu, Guo, Liang, Ouyang, et~al.]{li2025triposg}
Yangguang Li, Zi-Xin Zou, Zexiang Liu, Dehu Wang, Yuan Liang, Zhipeng Yu, Xingchao Liu, Yuan-Chen Guo, Ding Liang, Wanli Ouyang, et~al.
\newblock Triposg: High-fidelity 3d shape synthesis using large-scale rectified flow models.
\newblock \emph{arXiv preprint arXiv:2502.06608}, 2025.

\bibitem[Lin et~al.(2023)Lin, Gao, Tang, Takikawa, Zeng, Huang, Kreis, Fidler, Liu, and Lin]{lin2023magic3d}
Chen-Hsuan Lin, Jun Gao, Luming Tang, Towaki Takikawa, Xiaohui Zeng, Xun Huang, Karsten Kreis, Sanja Fidler, Ming-Yu Liu, and Tsung-Yi Lin.
\newblock Magic3d: High-resolution text-to-3d content creation.
\newblock In \emph{Proceedings of the IEEE/CVF Conference on Computer Vision and Pattern Recognition}, pages 300--309, 2023.

\bibitem[Liu et~al.(2024{\natexlab{a}})Liu, Huang, Huang, Chen, Hou, Tang, Liu, Ouyang, Zuo, Jiang, et~al.]{liu2024comprehensive}
Jian Liu, Xiaoshui Huang, Tianyu Huang, Lu~Chen, Yuenan Hou, Shixiang Tang, Ziwei Liu, Wanli Ouyang, Wangmeng Zuo, Junjun Jiang, et~al.
\newblock A comprehensive survey on 3d content generation.
\newblock \emph{arXiv preprint arXiv:2402.01166}, 2024{\natexlab{a}}.

\bibitem[Liu et~al.(2024{\natexlab{b}})Liu, Tam, Mahdavi-Amiri, and Savva]{liu2024cage}
Jiayi Liu, Hou In~Ivan Tam, Ali Mahdavi-Amiri, and Manolis Savva.
\newblock Cage: Controllable articulation generation.
\newblock In \emph{Proceedings of the IEEE/CVF Conference on Computer Vision and Pattern Recognition}, pages 17880--17889, 2024{\natexlab{b}}.

\bibitem[Liu et~al.(2020)Liu, Gu, Zaw~Lin, Chua, and Theobalt]{liu2020neural}
Lingjie Liu, Jiatao Gu, Kyaw Zaw~Lin, Tat-Seng Chua, and Christian Theobalt.
\newblock Neural sparse voxel fields.
\newblock \emph{Advances in Neural Information Processing Systems}, 33:\penalty0 15651--15663, 2020.

\bibitem[Liu et~al.(2023{\natexlab{a}})Liu, Shi, Chen, Zhang, Xu, Wei, Chen, Zeng, Gu, and Su]{liu2023onepp}
Minghua Liu, Ruoxi Shi, Linghao Chen, Zhuoyang Zhang, Chao Xu, Xinyue Wei, Hansheng Chen, Chong Zeng, Jiayuan Gu, and Hao Su.
\newblock One-2-3-45++: Fast single image to 3d objects with consistent multi-view generation and 3d diffusion.
\newblock \emph{arXiv preprint arXiv:2311.07885}, 2023{\natexlab{a}}.

\bibitem[Liu et~al.(2024{\natexlab{c}})Liu, Xu, Jin, Chen, Varma~T, Xu, and Su]{liu2024one}
Minghua Liu, Chao Xu, Haian Jin, Linghao Chen, Mukund Varma~T, Zexiang Xu, and Hao Su.
\newblock One-2-3-45: Any single image to 3d mesh in 45 seconds without per-shape optimization.
\newblock \emph{Advances in Neural Information Processing Systems}, 36, 2024{\natexlab{c}}.

\bibitem[Liu et~al.(2023{\natexlab{b}})Liu, Wu, Van~Hoorick, Tokmakov, Zakharov, and Vondrick]{liu2023zero}
Ruoshi Liu, Rundi Wu, Basile Van~Hoorick, Pavel Tokmakov, Sergey Zakharov, and Carl Vondrick.
\newblock Zero-1-to-3: Zero-shot one image to 3d object.
\newblock In \emph{Proceedings of the IEEE/CVF international conference on computer vision}, pages 9298--9309, 2023{\natexlab{b}}.

\bibitem[Long et~al.(2024)Long, Guo, Lin, Liu, Dou, Liu, Ma, Zhang, Habermann, Theobalt, and Wang]{long2024wonder3d}
Xiaoxiao Long, Yuan{-}Chen Guo, Cheng Lin, Yuan Liu, Zhiyang Dou, Lingjie Liu, Yuexin Ma, Song{-}Hai Zhang, Marc Habermann, Christian Theobalt, and Wenping Wang.
\newblock Wonder3d: Single image to 3d using cross-domain diffusion.
\newblock In \emph{{CVPR}}, pages 9970--9980. {IEEE}, 2024.

\bibitem[Lorensen and Cline(1987)]{lorensen1987marching}
William~E. Lorensen and Harvey~E. Cline.
\newblock Marching cubes: {A} high resolution 3d surface construction algorithm.
\newblock In \emph{{SIGGRAPH}}, pages 163--169. {ACM}, 1987.

\bibitem[Luo et~al.(2021)Luo, Li, Zhang, and Lee]{luo2021surfgen}
Andrew Luo, Tianqin Li, Wen-Hao Zhang, and Tai~Sing Lee.
\newblock Surfgen: Adversarial 3d shape synthesis with explicit surface discriminators.
\newblock In \emph{Proceedings of the IEEE/CVF International Conference on Computer Vision}, pages 16238--16248, 2021.

\bibitem[Luo and Hu(2021)]{luo2021diffusion}
Shitong Luo and Wei Hu.
\newblock Diffusion probabilistic models for 3d point cloud generation.
\newblock In \emph{Proceedings of the IEEE/CVF Conference on Computer Vision and Pattern Recognition}, pages 2837--2845, 2021.

\bibitem[Lyu et~al.(2021)Lyu, Kong, Xu, Pan, and Lin]{lyu2021conditional}
Zhaoyang Lyu, Zhifeng Kong, Xudong Xu, Liang Pan, and Dahua Lin.
\newblock A conditional point diffusion-refinement paradigm for 3d point cloud completion.
\newblock \emph{arXiv preprint arXiv:2112.03530}, 2021.

\bibitem[Maturana and Scherer(2015)]{maturana2015voxnet}
Daniel Maturana and Sebastian Scherer.
\newblock Voxnet: A 3d convolutional neural network for real-time object recognition.
\newblock In \emph{2015 IEEE/RSJ international conference on intelligent robots and systems (IROS)}, pages 922--928. IEEE, 2015.

\bibitem[Mescheder et~al.(2019)Mescheder, Oechsle, Niemeyer, Nowozin, and Geiger]{mescheder2019occupancy}
Lars Mescheder, Michael Oechsle, Michael Niemeyer, Sebastian Nowozin, and Andreas Geiger.
\newblock Occupancy networks: Learning 3d reconstruction in function space.
\newblock In \emph{Proceedings of the IEEE/CVF conference on computer vision and pattern recognition}, pages 4460--4470, 2019.

\bibitem[Mildenhall et~al.(2021)Mildenhall, Srinivasan, Tancik, Barron, Ramamoorthi, and Ng]{mildenhall2021nerf}
Ben Mildenhall, Pratul~P Srinivasan, Matthew Tancik, Jonathan~T Barron, Ravi Ramamoorthi, and Ren Ng.
\newblock Nerf: Representing scenes as neural radiance fields for view synthesis.
\newblock \emph{Communications of the ACM}, 65\penalty0 (1):\penalty0 99--106, 2021.

\bibitem[Mittal et~al.(2022)Mittal, Cheng, Singh, and Tulsiani]{mittal2022autosdf}
Paritosh Mittal, Yen-Chi Cheng, Maneesh Singh, and Shubham Tulsiani.
\newblock Autosdf: Shape priors for 3d completion, reconstruction and generation.
\newblock In \emph{Proceedings of the IEEE/CVF Conference on Computer Vision and Pattern Recognition}, pages 306--315, 2022.

\bibitem[M{\"u}ller et~al.(2022)M{\"u}ller, Evans, Schied, and Keller]{muller2022instant}
Thomas M{\"u}ller, Alex Evans, Christoph Schied, and Alexander Keller.
\newblock Instant neural graphics primitives with a multiresolution hash encoding.
\newblock \emph{ACM transactions on graphics (TOG)}, 41\penalty0 (4):\penalty0 1--15, 2022.

\bibitem[Nam et~al.(2022)Nam, Khlifi, Rodriguez, Tono, Zhou, and Guerrero]{nam20223d}
Gimin Nam, Mariem Khlifi, Andrew Rodriguez, Alberto Tono, Linqi Zhou, and Paul Guerrero.
\newblock 3d-ldm: Neural implicit 3d shape generation with latent diffusion models.
\newblock \emph{arXiv preprint arXiv:2212.00842}, 2022.

\bibitem[Nash et~al.(2020)Nash, Ganin, Eslami, and Battaglia]{nash2020polygen}
Charlie Nash, Yaroslav Ganin, SM~Ali Eslami, and Peter Battaglia.
\newblock Polygen: An autoregressive generative model of 3d meshes.
\newblock In \emph{International conference on machine learning}, pages 7220--7229. PMLR, 2020.

\bibitem[Nguyen-Phuoc et~al.(2020)Nguyen-Phuoc, Richardt, Mai, Yang, and Mitra]{nguyen2020blockgan}
Thu~H Nguyen-Phuoc, Christian Richardt, Long Mai, Yongliang Yang, and Niloy Mitra.
\newblock Blockgan: Learning 3d object-aware scene representations from unlabelled images.
\newblock \emph{Advances in neural information processing systems}, 33:\penalty0 6767--6778, 2020.

\bibitem[Nichol et~al.(2022)Nichol, Jun, Dhariwal, Mishkin, and Chen]{nichol2022point}
Alex Nichol, Heewoo Jun, Prafulla Dhariwal, Pamela Mishkin, and Mark Chen.
\newblock Point-e: A system for generating 3d point clouds from complex prompts.
\newblock \emph{arXiv preprint arXiv:2212.08751}, 2022.

\bibitem[Park et~al.(2019)Park, Florence, Straub, Newcombe, and Lovegrove]{park2019deepsdf}
Jeong~Joon Park, Peter Florence, Julian Straub, Richard Newcombe, and Steven Lovegrove.
\newblock Deepsdf: Learning continuous signed distance functions for shape representation.
\newblock In \emph{Proceedings of the IEEE/CVF conference on computer vision and pattern recognition}, pages 165--174, 2019.

\bibitem[Peebles and Xie(2023)]{peebles2023scalable}
William Peebles and Saining Xie.
\newblock Scalable diffusion models with transformers.
\newblock In \emph{Proceedings of the IEEE/CVF International Conference on Computer Vision}, pages 4195--4205, 2023.

\bibitem[Peng et~al.(2020)Peng, Niemeyer, Mescheder, Pollefeys, and Geiger]{peng2020convolutional}
Songyou Peng, Michael Niemeyer, Lars Mescheder, Marc Pollefeys, and Andreas Geiger.
\newblock Convolutional occupancy networks.
\newblock In \emph{Computer Vision--ECCV 2020: 16th European Conference, Glasgow, UK, August 23--28, 2020, Proceedings, Part III 16}, pages 523--540. Springer, 2020.

\bibitem[Po et~al.(2024)Po, Yifan, Golyanik, Aberman, Barron, Bermano, Chan, Dekel, Holynski, Kanazawa, et~al.]{po2024state}
Ryan Po, Wang Yifan, Vladislav Golyanik, Kfir Aberman, Jonathan~T Barron, Amit Bermano, Eric Chan, Tali Dekel, Aleksander Holynski, Angjoo Kanazawa, et~al.
\newblock State of the art on diffusion models for visual computing.
\newblock In \emph{Computer graphics forum}, volume~43, page e15063. Wiley Online Library, 2024.

\bibitem[Poole et~al.(2022)Poole, Jain, Barron, and Mildenhall]{poole2022dreamfusion}
Ben Poole, Ajay Jain, Jonathan~T Barron, and Ben Mildenhall.
\newblock Dreamfusion: Text-to-3d using 2d diffusion.
\newblock \emph{arXiv preprint arXiv:2209.14988}, 2022.

\bibitem[Ren et~al.(2024)Ren, Huang, Zeng, Museth, Fidler, and Williams]{ren2024xcube}
Xuanchi Ren, Jiahui Huang, Xiaohui Zeng, Ken Museth, Sanja Fidler, and Francis Williams.
\newblock Xcube: Large-scale 3d generative modeling using sparse voxel hierarchies.
\newblock In \emph{Proceedings of the IEEE/CVF Conference on Computer Vision and Pattern Recognition}, pages 4209--4219, 2024.

\bibitem[Rombach et~al.(2022)Rombach, Blattmann, Lorenz, Esser, and Ommer]{rombach2022high}
Robin Rombach, Andreas Blattmann, Dominik Lorenz, Patrick Esser, and Bj{\"o}rn Ommer.
\newblock High-resolution image synthesis with latent diffusion models.
\newblock In \emph{Proceedings of the IEEE/CVF conference on computer vision and pattern recognition}, pages 10684--10695, 2022.

\bibitem[Shi et~al.(2023{\natexlab{a}})Shi, Chen, Zhang, Liu, Xu, Wei, Chen, Zeng, and Su]{shi2023zero123++}
Ruoxi Shi, Hansheng Chen, Zhuoyang Zhang, Minghua Liu, Chao Xu, Xinyue Wei, Linghao Chen, Chong Zeng, and Hao Su.
\newblock Zero123++: a single image to consistent multi-view diffusion base model.
\newblock \emph{arXiv preprint arXiv:2310.15110}, 2023{\natexlab{a}}.

\bibitem[Shi et~al.(2023{\natexlab{b}})Shi, Wang, Ye, Long, Li, and Yang]{shi2023mvdream}
Yichun Shi, Peng Wang, Jianglong Ye, Mai Long, Kejie Li, and Xiao Yang.
\newblock Mvdream: Multi-view diffusion for 3d generation.
\newblock \emph{arXiv preprint arXiv:2308.16512}, 2023{\natexlab{b}}.

\bibitem[Shu et~al.(2019)Shu, Park, and Kwon]{shu20193d}
Dong~Wook Shu, Sung~Woo Park, and Junseok Kwon.
\newblock 3d point cloud generative adversarial network based on tree structured graph convolutions.
\newblock In \emph{Proceedings of the IEEE/CVF international conference on computer vision}, pages 3859--3868, 2019.

\bibitem[Siddiqui et~al.(2023)Siddiqui, Alliegro, Artemov, Tommasi, Sirigatti, Rosov, Dai, and Nie{\ss}ner]{siddiqui2023meshgpt}
Yawar Siddiqui, Antonio Alliegro, Alexey Artemov, Tatiana Tommasi, Daniele Sirigatti, Vladislav Rosov, Angela Dai, and Matthias Nie{\ss}ner.
\newblock Meshgpt: Generating triangle meshes with decoder-only transformers.
\newblock \emph{arXiv preprint arXiv:2311.15475}, 2023.

\bibitem[Sohl-Dickstein et~al.(2015)Sohl-Dickstein, Weiss, Maheswaranathan, and Ganguli]{sohl2015deep}
Jascha Sohl-Dickstein, Eric Weiss, Niru Maheswaranathan, and Surya Ganguli.
\newblock Deep unsupervised learning using nonequilibrium thermodynamics.
\newblock In \emph{International conference on machine learning}, pages 2256--2265. PMLR, 2015.

\bibitem[Sun et~al.(2022)Sun, Sun, and Chen]{sun2022direct}
Cheng Sun, Min Sun, and Hwann-Tzong Chen.
\newblock Direct voxel grid optimization: Super-fast convergence for radiance fields reconstruction.
\newblock In \emph{Proceedings of the IEEE/CVF conference on computer vision and pattern recognition}, pages 5459--5469, 2022.

\bibitem[Sun et~al.(2023)Sun, Zhang, Shao, Wang, Liu, Xie, and Liu]{sun2023dreamcraft3d}
Jingxiang Sun, Bo~Zhang, Ruizhi Shao, Lizhen Wang, Wen Liu, Zhenda Xie, and Yebin Liu.
\newblock Dreamcraft3d: Hierarchical 3d generation with bootstrapped diffusion prior.
\newblock \emph{arXiv preprint arXiv:2310.16818}, 2023.

\bibitem[Sun et~al.(2020)Sun, Wang, Liu, Siegel, and Sarma]{sun2020pointgrow}
Yongbin Sun, Yue Wang, Ziwei Liu, Joshua Siegel, and Sanjay Sarma.
\newblock Pointgrow: Autoregressively learned point cloud generation with self-attention.
\newblock In \emph{Proceedings of the IEEE/CVF Winter Conference on Applications of Computer Vision}, pages 61--70, 2020.

\bibitem[Tang et~al.(2023)Tang, Ren, Zhou, Liu, and Zeng]{tang2023dreamgaussian}
Jiaxiang Tang, Jiawei Ren, Hang Zhou, Ziwei Liu, and Gang Zeng.
\newblock Dreamgaussian: Generative gaussian splatting for efficient 3d content creation.
\newblock \emph{arXiv preprint arXiv:2309.16653}, 2023.

\bibitem[Wang et~al.(2024{\natexlab{a}})Wang, Peng, Liu, Gu, and Hu]{wangdiffusion}
Chen Wang, Hao-Yang Peng, Ying-Tian Liu, Jiatao Gu, and Shi-Min Hu.
\newblock Diffusion models for 3d generation: A survey.
\newblock \emph{arXiv}, 2024{\natexlab{a}}.

\bibitem[Wang et~al.(2022{\natexlab{a}})Wang, Liu, and Tong]{wang2022doctree}
Peng-Shuai Wang, Yang Liu, and Xin Tong.
\newblock Dual octree graph networks for learning adaptive volumetric shape representations.
\newblock \emph{ACM SIGGRAPH 2022 Conference Papers}, 2022{\natexlab{a}}.

\bibitem[Wang et~al.(2022{\natexlab{b}})Wang, Liu, and Tong]{wang2022dual}
Peng-Shuai Wang, Yang Liu, and Xin Tong.
\newblock Dual octree graph networks for learning adaptive volumetric shape representations.
\newblock \emph{ACM Transactions on Graphics (TOG)}, 41\penalty0 (4):\penalty0 1--15, 2022{\natexlab{b}}.

\bibitem[Wang et~al.(2023)Wang, Zhang, Zhang, Gu, Bao, Baltrusaitis, Shen, Chen, Wen, Chen, et~al.]{wang2023rodin}
Tengfei Wang, Bo~Zhang, Ting Zhang, Shuyang Gu, Jianmin Bao, Tadas Baltrusaitis, Jingjing Shen, Dong Chen, Fang Wen, Qifeng Chen, et~al.
\newblock Rodin: A generative model for sculpting 3d digital avatars using diffusion.
\newblock In \emph{Proceedings of the IEEE/CVF conference on computer vision and pattern recognition}, pages 4563--4573, 2023.

\bibitem[Wang et~al.(2024{\natexlab{b}})Wang, Lu, Wang, Bao, Li, Su, and Zhu]{wang2024prolificdreamer}
Zhengyi Wang, Cheng Lu, Yikai Wang, Fan Bao, Chongxuan Li, Hang Su, and Jun Zhu.
\newblock Prolificdreamer: High-fidelity and diverse text-to-3d generation with variational score distillation.
\newblock \emph{Advances in Neural Information Processing Systems}, 36, 2024{\natexlab{b}}.

\bibitem[Wizadwongsa et~al.(2024)Wizadwongsa, Zhou, Li, and Park]{wizadwongsa2024taming}
Suttisak Wizadwongsa, Jinfan Zhou, Edward Li, and Jeong~Joon Park.
\newblock Taming feed-forward reconstruction models as latent encoders for 3d generative models.
\newblock \emph{arXiv preprint arXiv:2501.00651}, 2024.

\bibitem[Wu et~al.(2016)Wu, Zhang, Xue, Freeman, and Tenenbaum]{wu2016learning}
Jiajun Wu, Chengkai Zhang, Tianfan Xue, Bill Freeman, and Josh Tenenbaum.
\newblock Learning a probabilistic latent space of object shapes via 3d generative-adversarial modeling.
\newblock \emph{Advances in neural information processing systems}, 29, 2016.

\bibitem[Wu et~al.(2020)Wu, Zhuang, Xu, Zhang, and Chen]{wu2020pq}
Rundi Wu, Yixin Zhuang, Kai Xu, Hao Zhang, and Baoquan Chen.
\newblock Pq-net: A generative part seq2seq network for 3d shapes.
\newblock In \emph{Proceedings of the IEEE/CVF Conference on Computer Vision and Pattern Recognition}, pages 829--838, 2020.

\bibitem[Wu et~al.(2024{\natexlab{a}})Wu, Lin, Zhang, Zeng, Xu, Torr, Cao, and Yao]{wu2024direct3d}
Shuang Wu, Youtian Lin, Feihu Zhang, Yifei Zeng, Jingxi Xu, Philip Torr, Xun Cao, and Yao Yao.
\newblock Direct3d: Scalable image-to-3d generation via 3d latent diffusion transformer.
\newblock \emph{arXiv preprint arXiv:2405.14832}, 2024{\natexlab{a}}.

\bibitem[Wu et~al.(2024{\natexlab{b}})Wu, Li, Yan, Shang, Sun, Wang, Cui, Liu, Sato, Li, et~al.]{wu2024blockfusion}
Zhennan Wu, Yang Li, Han Yan, Taizhang Shang, Weixuan Sun, Senbo Wang, Ruikai Cui, Weizhe Liu, Hiroyuki Sato, Hongdong Li, et~al.
\newblock Blockfusion: Expandable 3d scene generation using latent tri-plane extrapolation.
\newblock \emph{ACM Transactions on Graphics (TOG)}, 43\penalty0 (4):\penalty0 1--17, 2024{\natexlab{b}}.

\bibitem[Wu et~al.(2015)Wu, Song, Khosla, Yu, Zhang, Tang, and Xiao]{wu20153d}
Zhirong Wu, Shuran Song, Aditya Khosla, Fisher Yu, Linguang Zhang, Xiaoou Tang, and Jianxiong Xiao.
\newblock 3d shapenets: A deep representation for volumetric shapes.
\newblock In \emph{Proceedings of the IEEE conference on computer vision and pattern recognition}, pages 1912--1920, 2015.

\bibitem[Xiang et~al.(2024)Xiang, Lv, Xu, Deng, Wang, Zhang, Chen, Tong, and Yang]{xiang2024structured}
Jianfeng Xiang, Zelong Lv, Sicheng Xu, Yu~Deng, Ruicheng Wang, Bowen Zhang, Dong Chen, Xin Tong, and Jiaolong Yang.
\newblock Structured 3d latents for scalable and versatile 3d generation.
\newblock \emph{arXiv preprint arXiv:2412.01506}, 2024.

\bibitem[Xie et~al.(2024)Xie, Zheng, Huang, Chen, Wang, Ye, Chen, and Huo]{xie2024ldm}
Rengan Xie, Wenting Zheng, Kai Huang, Yizheng Chen, Qi~Wang, Qi~Ye, Wei Chen, and Yuchi Huo.
\newblock Ldm: Large tensorial sdf model for textured mesh generation.
\newblock \emph{arXiv preprint arXiv:2405.14580}, 2024.

\bibitem[Xu et~al.(2019)Xu, Wang, Ceylan, Mech, and Neumann]{xu2019disn}
Qiangeng Xu, Weiyue Wang, Duygu Ceylan, Radomir Mech, and Ulrich Neumann.
\newblock Disn: Deep implicit surface network for high-quality single-view 3d reconstruction.
\newblock \emph{Advances in neural information processing systems}, 32, 2019.

\bibitem[Yan et~al.(2022)Yan, Lin, Mitra, Lischinski, Cohen-Or, and Huang]{yan2022shapeformer}
Xingguang Yan, Liqiang Lin, Niloy~J Mitra, Dani Lischinski, Daniel Cohen-Or, and Hui Huang.
\newblock Shapeformer: Transformer-based shape completion via sparse representation.
\newblock In \emph{Proceedings of the IEEE/CVF Conference on Computer Vision and Pattern Recognition}, pages 6239--6249, 2022.

\bibitem[Yang et~al.(2019)Yang, Huang, Hao, Liu, Belongie, and Hariharan]{yang2019pointflow}
Guandao Yang, Xun Huang, Zekun Hao, Ming-Yu Liu, Serge Belongie, and Bharath Hariharan.
\newblock Pointflow: 3d point cloud generation with continuous normalizing flows.
\newblock In \emph{Proceedings of the IEEE/CVF international conference on computer vision}, pages 4541--4550, 2019.

\bibitem[Yang et~al.(2024)Yang, Shi, Zhang, Yang, Wang, Zhao, Liu, Wang, Lin, Yu, et~al.]{yang2024hunyuan3d}
Xianghui Yang, Huiwen Shi, Bowen Zhang, Fan Yang, Jiacheng Wang, Hongxu Zhao, Xinhai Liu, Xinzhou Wang, Qingxiang Lin, Jiaao Yu, et~al.
\newblock Hunyuan3d-1.0: A unified framework for text-to-3d and image-to-3d generation.
\newblock \emph{arXiv preprint arXiv:2411.02293}, 2024.

\bibitem[Zamorski et~al.(2020)Zamorski, Zikeba, Klukowski, Nowak, Kurach, Stokowiec, and Trzci{\'n}ski]{zamorski2020adversarial}
Maciej Zamorski, Maciej Zikeba, Piotr Klukowski, Rafa{\l} Nowak, Karol Kurach, Wojciech Stokowiec, and Tomasz Trzci{\'n}ski.
\newblock Adversarial autoencoders for compact representations of 3d point clouds.
\newblock \emph{Computer Vision and Image Understanding}, 193:\penalty0 102921, 2020.

\bibitem[Zhang et~al.(2023)Zhang, Tang, Niessner, and Wonka]{zhang20233dshape2vecset}
Biao Zhang, Jiapeng Tang, Matthias Niessner, and Peter Wonka.
\newblock 3dshape2vecset: A 3d shape representation for neural fields and generative diffusion models.
\newblock \emph{ACM Transactions on Graphics (TOG)}, 42\penalty0 (4):\penalty0 1--16, 2023.

\bibitem[Zhang et~al.(2024)Zhang, Wang, Zhang, Qiu, Pang, Jiang, Yang, Xu, and Yu]{zhang2024clay}
Longwen Zhang, Ziyu Wang, Qixuan Zhang, Qiwei Qiu, Anqi Pang, Haoran Jiang, Wei Yang, Lan Xu, and Jingyi Yu.
\newblock Clay: A controllable large-scale generative model for creating high-quality 3d assets.
\newblock \emph{arXiv preprint arXiv:2406.13897}, 2024.

\bibitem[Zhao and Larsen(2024)]{zhao2024challenges}
Ke~Zhao and Andreas Larsen.
\newblock Challenges and opportunities in 3d content generation.
\newblock \emph{arXiv preprint arXiv:2405.15335}, 2024.

\bibitem[Zhao et~al.(2023)Zhao, Liu, Chen, Zeng, Wang, Cheng, Fu, Chen, Yu, and Gao]{zhao2023michelangelo}
Zibo Zhao, Wen Liu, Xin Chen, Xianfang Zeng, Rui Wang, Pei Cheng, Bin Fu, Tao Chen, Gang Yu, and Shenghua Gao.
\newblock Michelangelo: Conditional 3d shape generation based on shape-image-text aligned latent representation.
\newblock In \emph{NeurIPS}, 2023.

\bibitem[Zheng et~al.(2023)Zheng, Pan, Wang, Tong, Liu, and Shum]{zheng2023locally}
Xin-Yang Zheng, Hao Pan, Peng-Shuai Wang, Xin Tong, Yang Liu, and Heung-Yeung Shum.
\newblock Locally attentional sdf diffusion for controllable 3d shape generation.
\newblock \emph{arXiv preprint arXiv:2305.04461}, 2023.

\bibitem[Zhou et~al.(2023)Zhou, Wang, Ma, Liu, Huang, and Wang]{zhou2023uni3d}
Junsheng Zhou, Jinsheng Wang, Baorui Ma, Yu-Shen Liu, Tiejun Huang, and Xinlong Wang.
\newblock Uni3d: Exploring unified 3d representation at scale.
\newblock \emph{arXiv preprint arXiv:2310.06773}, 2023.

\bibitem[Zhou et~al.(2021)Zhou, Du, and Wu]{zhou20213d}
Linqi Zhou, Yilun Du, and Jiajun Wu.
\newblock 3d shape generation and completion through point-voxel diffusion.
\newblock In \emph{Proceedings of the IEEE/CVF International Conference on Computer Vision}, pages 5826--5835, 2021.

\end{thebibliography}
\bibliographystyle{plainnat}

\appendix
\section*{Appendix}

\section{Representations and Algorithms in 3D Generative Modeling} \label{sec:representations}

Meshes are foundational in 3D computer graphics and vision, representing surfaces in terms of vertices, edges, and faces to form polygonal approximations of objects. While meshes are extensively used across various fields due to their broad compatibility with rendering pipelines and ability to depict complex geometries with high fidelity, processing them within neural network architectures presents significant challenges. The variable topology and connectivity inherent in meshes make them difficult to process using standard neural architectures that expect regular input structures. Consequently, processing meshes often requires specialized neural networks, such as graph neural networks~\citep{li2024advances}, or autoregressive architectures~\citep{yan2022shapeformer, mittal2022autosdf, siddiqui2023meshgpt} which can be complex and computationally inefficient. Moreover, high-resolution meshes can have substantial memory footprints, raising concerns about the scalability of the resulting learning pipelines. These limitations have led researchers to explore alternative three-dimensional representations derived from meshes through algorithmic conversions that typically do not require training to facilitate processing, learning, and rendering within deep neural frameworks \cite{li2024advances}.

\paragraph{Voxel grids} 
Voxel grids offer an intuitive and straightforward encoding for 3D objects analogous to pixel representations in 2D images. Early research predominantly relied on dense voxel grids~\citep{wu20153d, maturana2015voxnet, choy20163d, wu2016learning, brock2016generative, dai2018scancomplete}. Despite their convenience, voxel grids suffer from a large memory footprint, limiting the achievable resolution and computational efficiency.
Recent methods have revisited voxel grids to leverage their compatibility with convolutional neural networks (CNNs). For instance, X-Cube~\citep{ren2024xcube} introduces a multi-resolution approach involving denoising and decoding steps, while methods like LAS-Diffusion~\citep{zheng2023locally} and One-2-3-45++~\citep{liu2023onepp} utilize voxel grids in the initial stages of their generation pipelines. These approaches demonstrate that voxel grids can still be effective when combined with modern techniques like diffusion models.
To mitigate memory constraints, sparse or hierarchical voxel grids have been proposed~\citep{liu2020neural, fridovich2022plenoxels, sun2022direct}. Notably, Instant Neural Graphics Primitives (Instant-NGP)~\citep{muller2022instant} use a multi-level voxel grid encoded via a hash function, enabling fast optimization and rendering while maintaining a compact model size.

\paragraph{Implicit neural functions} 
Implicit representations model 3D geometry as continuous functions, allowing for high-resolution detail without the memory overhead of dense grids. Occupancy Networks~\citep{mescheder2019occupancy} introduced an implicit representation that predicts the occupancy probability of arbitrary points in 3D space, conditioned on input data such as images or point clouds. This approach allows for smooth and detailed surface representations. 
Extensions of this idea include Convolutional Occupancy Networks~\citep{peng2020convolutional} and Neural Implicit Surfaces~\citep{atzmon2020sal}, which improve the ability to capture fine geometric details. Recently, \citet{zhang20233dshape2vecset} proposed 3DShape2VecSet, employing a Transformer-based encoder and decoder where occupancy at query points is predicted via cross-attention mechanisms. This representation and follow-up work~\cite{chen2024dora} have been used for conditional generation, achieving remarkable results~\citep{zhang2024clay, yang2024hunyuan3d, li2025triposg}. 
In contrast to these works, Neural Radiance Fields (NeRFs) \citep{mildenhall2021nerf} jointly model geometry and texture using implicit representations, enabling rendering from arbitrary views via volume ray casting. NeRFs are especially popular for 3D generative methods that only rely on image data for supervision, such as DreamFusion \citep{poole2022dreamfusion}.

\paragraph{Signed Distance Functions (SDF)}
Signed Distance Functions (SDFs) provide a scalar field where each point in space is assigned a value representing its signed distance to the closest surface, with negative values indicating points inside the object. SDFs offer a more expressive alternative to occupancy functions, capturing both the geometry and topology of 3D shapes.
DeepSDF~\citep{park2019deepsdf} pioneered the use of neural networks to learn continuous SDF representations from data. Building on this, DISN~\citep{xu2019disn} introduced an amortized approach that eliminates the need for test-time optimization, enabling more efficient inference.
SDFs can be discretized over a voxel grid, resulting in a \textit{sampled SDF} or \textit{grid SDF}, which facilitates the use of convolutional architectures. SDFusion~\citep{cheng2023sdfusion} demonstrated the effectiveness of this approach for 3D generation. Additionally, \citet{mittal2022autosdf} proposed AutoSDF, which generates SDF grids for shape patches autoregressively after encoding them with a VQ-VAE.
Further advancements include methods like 3D-LDM~\citep{nam20223d}, where an MLP autoencoder predicts SDF values from latent codes and query points, blending occupancy and SDF representations. \citet{xie2024ldm} compared different representations, concluding that tensorial SDFs outperform triplane SDFs and tensorial NeRFs in terms of fidelity and efficiency.

\vspace{-3mm}
\paragraph{Triplanes}
Triplane representations encode 3D scenes using three orthogonal feature planes $(\boldsymbol{xy}, \boldsymbol{xz}, \boldsymbol{yz}) \in \mathbb{R}^{N \times N \times C}$, significantly reducing memory requirements while retaining spatial information. This approach was popularized by EG3D~\citep{chan2022efficient}, which employs a triplane representation within a Generative Adversarial Network (GAN) framework for high-quality 3D-aware image synthesis.
Recent works have extended triplane representations to 3D generative modeling. Approaches such as 3DGen \cite{gupta20233dgen}, RODIN~\cite{wang2023rodin}, Blockfusion \cite{wu2024blockfusion} and Direct3D \cite{wu2024direct3d} rely on latent diffusion to denoise and then up-sample triplane latents, which are then decoded via a lightweight MLP occupancy network and rendered volumetrically, facilitating the generation of high-fidelity 3D avatars or even complete 3D environments.
Practically, triplanes can be obtained from occupancy data by pretraining a dedicated PointNet-UNet-OccNet autoencoder on a reconstruction loss, as detailed in~\cite{peng2020convolutional}. In this approach, the triplane latent resolution can be adjusted using specific UNet layers within the autoencoder. Alternatively, a triplane dataset can be directly optimized from a mesh dataset, as proposed in~\cite{wu2024blockfusion}, by jointly training an MLP decoder and its triplane input—considered here as an optimization variable on par with the weights and biases of the MLP decoder—on a reconstruction loss over the entire dataset. In this context, since the triplane equivalent of each asset in the database is readily available after pretraining, the encoder part of the generative approach consists only of projection layers. These layers encode high-dimensional triplanes into lower-dimensional triplane latents, on which the denoising diffusion model will subsequently act.

Apart from reconstruction and generation tasks, 3D representations were leveraged for CLIP-like foundation models that learn an encoding of 3D objects, which is used for downstream tasks such as classification or segmentation~\cite{zhou2023uni3d, hegde2023clip}.

\section{Generative pipelines}\label{app:diffusion}

Diffusion models~\citep{sohl2015deep} have emerged as a powerful class of generative models, demonstrating remarkable success in image synthesis~\citep{ho2020denoising}. They operate by progressively corrupting training data through the sequential addition of Gaussian noise (forward process) and then learning to recover the original data by reversing this noising process (reverse process). This framework has been extended to 3D data, allowing for the generation of complex 3D structures~\citep{lyu2021conditional, zhou20213d}.

\vspace{-3mm}
\paragraph{Forward Process}
Given a data sample $\boldsymbol{x}_0$ drawn from a distribution $q\left(\boldsymbol{x}_0\right)$, the forward diffusion process generates a sequence of increasingly noisy samples $\boldsymbol{x}_t$, $\forall t \in \left[ 1, \cdots, T\right]$ from $\boldsymbol{x}_0$ by adding Gaussian noise $\boldsymbol{\varepsilon} \sim \mathcal{N}\left(\boldsymbol{0}, \boldsymbol{I}\right)$ at each timestep according to a predefined variance schedule $0 < \beta_1 < \cdots < \beta_T < 1$: 
\begin{eqnarray}
    q(\boldsymbol{x}_{1:T}|\boldsymbol{x}_0) &:=& \prod_{t = 1}^{T} q\left(\boldsymbol{x}_{t}|\boldsymbol{x}_{t-1}\right),\\  q\left(\boldsymbol{x}_{t}|\boldsymbol{x}_{t-1}\right) &:=& \mathcal{N}\left(\sqrt{1-\beta_{t}}\boldsymbol{x}_{t-1}, \beta_{t}\boldsymbol{I}\right).
\end{eqnarray}
A relevant property of such a diffusion process is that $\boldsymbol{x}_t$ can be sampled from $\boldsymbol{x}_0$ using the closed-form expression:
\begin{equation}
    q(\boldsymbol{x}_{t}|\boldsymbol{x}_0) = \mathcal{N}(\sqrt{\Bar{\alpha}_{t}}\boldsymbol{x}_{0}, \left(1-\Bar{\alpha}_{t}\right)\boldsymbol{I}),
\end{equation}
where $\alpha_{t} := 1-\beta_{t}$ and $\Bar{\alpha}_{t} = \prod_{s=1}^{t} \alpha_{s}$. This leads to the practical sampling equation: 
\begin{equation}
\boldsymbol{x}_t = \sqrt{\bar{\alpha}_t} \boldsymbol{x}_0 + \sqrt{1 - \bar{\alpha}_t} \boldsymbol{\varepsilon}, \quad \boldsymbol{\varepsilon} \sim \mathcal{N}(\boldsymbol{0}, \boldsymbol{I}).
\end{equation}

\vspace{-3mm}
\paragraph{Reverse Process.}
Initiating from a standard Gaussian distribution, $\boldsymbol{x}_T \sim \mathcal{N}(0,\boldsymbol{I})$, a denoising model $p_\theta$ parameterized by trainable weights $\theta$, learns to approximate a series of Gaussian transitions $p_\theta\left(\boldsymbol{x}_{t-1}|\boldsymbol{x}_t\right)$. These transitions incrementally denoise the signal such that
\vspace{-2mm}
\begin{eqnarray}
    p_\theta\left(\boldsymbol{x}_{0:T}\right) &:=& p_\theta\left(\boldsymbol{x}_{T}\right) \prod_{t = 1}^{T} p_\theta\left(\boldsymbol{x}_{t-1}|\boldsymbol{x}_{t}\right), \\
    p_\theta\left(\boldsymbol{x}_{t-1}|\boldsymbol{x}_{t}\right) &:=& \mathcal{N}\left(\boldsymbol{\mu}_\theta\left(\boldsymbol{x}_t, t\right), \boldsymbol{\Sigma}_\theta\left(\boldsymbol{x}_t, t\right) \right).
\end{eqnarray}
Following the approach from \citep{ho2020denoising}, we define 
\begin{eqnarray}
    \boldsymbol{\mu}_\theta &=& \frac{1}{\sqrt{\alpha_{t}}}\left( \boldsymbol{x}_t - \frac{\beta_t}{\sqrt{1-\Bar{\alpha}_{t}}} \boldsymbol{\varepsilon}_\theta\left(\boldsymbol{x}_t, t\right)\right)\\
    \boldsymbol{\Sigma}_\theta(\boldsymbol{x}_t, t) &=& \sigma_t^2\boldsymbol{I}
\end{eqnarray}
yielding the following Langevin dynamics 
\begin{equation}
    \boldsymbol{x}_{t-1} = \frac{1}{\sqrt{\alpha_{t}}}\left( \boldsymbol{x}_t - \frac{\beta_t}{\sqrt{1-\Bar{\alpha}_{t}}} \boldsymbol{\varepsilon}_\theta\left(\boldsymbol{x}_t, t\right)\right) + \sigma_t \boldsymbol{z},
\end{equation}
where $\boldsymbol{z}\sim\mathcal{N}\left(\boldsymbol{0}, \boldsymbol{I}\right)$ and $\boldsymbol{\varepsilon}_\theta\left(\boldsymbol{x}_t, t\right)$ is a learnable network approximating the per-step noise on $\boldsymbol{x}_t$. 
\vspace{-3mm}
\paragraph{Loss Function}
The model is trained by minimizing the variational bound on negative log-likelihood. However, \citet{ho2020denoising} showed that a simplified loss focusing on the noise prediction yields good empirical results:
\begin{equation}
    \mathcal{L} = \mathbb{E}_{\boldsymbol{x}_0, \boldsymbol{\varepsilon}_t} \left[  \left\| \boldsymbol{\varepsilon}_t - \boldsymbol{\varepsilon}_\theta\left(\sqrt{\alpha_{t}}\boldsymbol{x}_0 + \sqrt{1 - \Bar{\alpha}_t}\boldsymbol{\varepsilon}_t,t\right)  \right\|^2 \right]
\end{equation}

This loss encourages the network to predict the noise added at each timestep, facilitating the denoising process during generation.

We adopt open-source implementations for DiT\footnote{\url{https://github.com/facebookresearch/DiT}} and a 3D U-Net\footnote{\url{https://github.com/CompVis/latent-diffusion/}}. Specific parameter settings will be provided as \texttt{hydra} config files with our open-source codebase.

\section{Evaluation Metrics}\label{app:metrics}

We use the following metrics for evaluating our approach.
The symmetric Chamfer distance is selected to measure the distance between two point clouds $X$ and $Y$.
\begin{equation}
    \mathrm{CD}(X,Y) = \frac{1}{|X|}\sum_{\rvx\in X} \min_{\rvy\in Y}\Vert \rvx - \rvy \Vert_2 + \frac{1}{|Y|}\sum_{\rvy\in Y} \min_{\rvx\in X}\Vert \rvy - \rvx \Vert_2.
    \label{eq:cd}
\end{equation}

The F-score is the harmonic mean of precision and recall for a generated mesh $\gG$ and a reference mesh $\gR$. The precision is defined as
\begin{equation}
    P(\tau) = \frac{100}{|\gG|} \sum_{\rvg \in \gG} \left[ \min_{\rvr \in \gR} \left\Vert \rvg-\rvr \right\Vert_2 < \tau \right],
\end{equation}
with $\rvg$ as points sampled on the surface of the generated mesh, $\rvr$ as points sampled from the reference mesh, $\tau$ as a threshold, and $[\cdot]$ as the Iversion bracket.
The recall is defined accordingly as 
\begin{equation}
    R(\tau) = \frac{100}{|\gR|} \sum_{\rvr \in \gR} \left[ \min_{\rvg \in \gG} \left\Vert \rvr-\rvg \right\Vert_2 < \tau \right].
\end{equation}
The final F-score is then computed as
\begin{equation}
    F(\tau) = \frac{2 P(\tau) R(\tau)}{P(\tau)+R(\tau)}.
    \label{eq:fscore}
\end{equation}

We define the normal consistency between a generated mesh $\gG$ and a reference mesh $\gR$ as
\begin{equation}
\mathrm{NC}(\gG,\gR) = \frac{1}{|\gG|} \sum_{\rvg \in \gG} \left< \rvn(\rvg), \rvn(N_\rvr) \right>.
\end{equation}
The function $\rvn(\cdot)$ retrieves the normal of the respective point, $\rvg$ is a point sampled on the surface of the generated mesh and $N_\rvr$ is the closest point on $\gR$ to the point $\rvg$. $\left<\cdot,\cdot\right>$ is the dot product which measures the similarity of the normals.

To measure the quality of the generated meshes w.r.t. a set of reference meshes we use the metrics Coverage ($\mathrm{COV}$), Minimum matching distance ($\mathrm{MMD}$), and 1-nearest neighbor accuracy ($\text{1-NNA}$).

\begin{equation}
    \mathrm{COV}(S_g, S_r) = \frac{|\{\mathrm{arg\,min}_{Y\in S_r} D(X,Y) | X\in S_g\}|}{|S_r|}.
    \label{eq:cov}
\end{equation}

\begin{equation}
    \mathrm{MMD}(S_g, S_r) = \frac{1}{|S_r|} \sum_{Y\in S_r}\underset{X\in S_g}{\mathrm{min}} D(X,Y).
    \label{eq:mmd}
\end{equation}

\begin{equation}
    \text{1-NNA}(S_g, S_r) = \frac{\sum_{X\in S_g}\mathbb{I}[N_x\in S_g] + \sum_{Y\in S_r} \mathbb{I}[N_y \in S_r]}{|S_g|+|S_r|}.
    \label{eq:1nna}
\end{equation}
$N_X$ is the nearest neighbor of $X$ in the set $S_r \cup S_g - \{X\}$. $S_r$, $S_g$ are the sets of reference and generated meshes and are of equal size.
We use the Chamfer distance for $D$ in all our experiments.

\section{Training Details}
All training parameters can be found in the config files in the \texttt{configs} directory in the root of the repository.
The main entry point is the \texttt{configs/train.yaml} files.
For specific models and experiments, training parameters can be found in the respective \texttt{configs/model} and \texttt{configs/experiment} subfolders. For instance, the configuration for the generation experiment with the SDF representation using an autoencoder and diffusion transformer on the Airplane category is in \texttt{configs/experiment/diffuse\_sdf\_ae\_dit\_airplane.yaml}. 

We give a summary of the most important training parameters here.
We trained all models using Adam as optimizer until convergence. The learning rate and schedule highly depends on the model used and have been found by hyperparameter search.
For most experiments, we use a step-wise exponential decay learning rate schedule. For Shape2VecSet and Triplane models, we use a half-cycle cosine decay started after a warm-up phase. 
We have trained the models using up to 2 NVIDIA A6000 GPUs. Training times for all models are within 5 days.

\section{Reconstruction and generation results on Objaverse data}\label{app:objaverse}

To validate our results on a second widely-used dataset, we repeat the analysis on Objaverse data. We selected five categories for which most category-labeled samples were available: chair, seashell, apple, banana, mug, shield and skateboard (together 876/113/183 train/val/test samples). We first train the encoders, then the diffusion models on the same data with the same split. 

The reconstruction results are shown in \autoref{tab:recon_objaverse}. These results mostly align with the ShapeNet results: 1) The value ranges are similar. 2) They confirm SDF AE as the best method for reconstruction. 3) The ranking of methods mostly aligns (compare \autoref{tab:reconstruction}). The only change for Objaverse is a better performance for DualOctree, surpassing Shape2vecset; however, these methods had similar F-score for ShapeNet (F-score of 79 for Shape2VecSet vs 76 for DualOctree). 

\begin{table}[h!]
\centering
\caption{Reconstruction performance of best generation approaches on the Objaverse dataset.}
\label{tab:recon_objaverse}
\resizebox{0.82\textwidth}{!}{
\begin{tabular}{lllll}
\toprule
\textbf{Representation} & \textbf{Encoder} &  \textbf{F-score $\uparrow$ (0.0125)} ($\tau=\frac{1}{80}$) & \textbf{CD} $\downarrow$ ($*1e-4$) & \textbf{NC} $\uparrow$ \\
\midrule
DualOctree & VAE &  83.86 $\pm$ 15.173 & 0.016 $\pm$ 10.51 & 0.857 $\pm$ 0.081 \\
SDF  & AE &  96.423 $\pm$ 6.107 & 0.009 $\pm$ 4.948 & 0.896 $\pm$ 0.062 \\
Shape2VecSet & AE & 67.578 $\pm$ 28.472 & 0.045 $\pm$ 55.053 & 0.815 $\pm$ 0.128 \\
Triplane & AE &  66.883 $\pm$ 21.589 & 0.036 $\pm$ 41.114 & 0.82 $\pm$ 0.092 \\
Voxel & AE &  96.878 $\pm$ 6.935 & 0.008 $\pm$ 7.339 & 0.85 $\pm$ 0.06 \\
\bottomrule
\end{tabular}
}
\end{table}

Furthermore, \autoref{tab:gen_objaverse} shows the metrics for unconditional generation. Only the Triplane results improved considerably compared to the experiments on ShapeNet, surpassing Shape2Vecset in the Objaverse experiments. The main result, putting DualOctree and SDF as the best methods, remains the same.  

It is worth noting that we computed the metrics using 183 samples for the generated and reference set respectively, since this is the size of the test set. As shown in \autoref{fig:shapenet_gt_uncond_metrics}, low sample size biases the results. Indeed, we see higher coverage and lower 1-NNA than for our evaluation on ShapeNet, where we evaluated on 400 samples.

\begin{table}[h!]
\centering
\caption{Evaluation metrics for unconditional generation trained on Objaverse dataset including COV, MMD, and 1-NNA scores.}
\label{tab:gen_objaverse}
\begin{tabular}{llllll}
\toprule
\textbf{Representation} & \textbf{Generator} & \textbf{Encoder} & \textbf{COV} $\uparrow$& \textbf{MMD} $\downarrow$& \textbf{1-NNA}$ \rightarrow$ 0.5 \\
\midrule
DualOctree & VAE &  UNet & 0.47541 & 0.0237217 & 0.584699 \\
SDF & AE &   DiT & 0.491803 & 0.0277514 & 0.68306 \\
Shape2VecSet & AE &  DiT & 0.349727 & 0.0298354 & 0.819672 \\
Triplane & AE &   UNet & 0.47541 & 0.0248811 & 0.729508 \\
Voxel & AE & DiT & 0.409836 & 0.0371019 & 0.803279 \\
\bottomrule
\end{tabular}
\end{table}

\section{Ablation study: Reconstruction performance}\label{app:ablation_recon}

\autoref{tab:ablationreconstruction} shows the results for reconstruction for the ShapeNet Chair category across all tested encoders. In most cases, a standard autoencoder (with LayerNorm) performs best. This is expected since penalizing the KL-divergence in VAEs stirs the distribution at the cost of lower reconstruction performance. However, we find that the encoder choice only plays a minor role compared to the differences between representations. 
Specifically, the inter-representation standard deviation (between best CD per representations) is 0.0098 whereas the intra-representation standard deviation (between encoder-wise CD for each representation) is only 0.0050\footnote{The intra-representation StD was only computed for the representations where results are available for all three encoder models}.

\begin{table}[ht]
\centering
\resizebox{0.7\linewidth}{!}{
\newcolumntype{H}{>{\setbox0=\hbox\bgroup}c<{\egroup}@{}}
\begin{tabular}{lllll}
\toprule
\textbf{Representation} & \textbf{Encoder} & \textbf{F-score $\uparrow$ ($\tau=\frac{1}{80}$)} & \textbf{CD $\downarrow$ (*1e-4)} & \textbf{NC} $\uparrow$ \\
\midrule
DualOctree  & AE & 91.629 \footnotesize{ $\pm$ 7.50} & 0.012 \footnotesize{ $\pm$ 0.01} & 0.821 \footnotesize{ $\pm$ 0.07} \\
DualOctree  & VAE & 83.152 \footnotesize{ $\pm$ 12.38} & 0.017 \footnotesize{ $\pm$ 0.01} & 0.798 \footnotesize{ $\pm$ 0.07} \\
DualOctree  & VQVAE & 73.763 \footnotesize{ $\pm$ 14.23} & 0.023 \footnotesize{ $\pm$ 0.01} & 0.773 \footnotesize{ $\pm$ 0.07} \\
\midrule
NeRF  & AE & 58.162 \footnotesize{ $\pm$ 13.74} & 0.032 \footnotesize{ $\pm$ 0.02} & 0.714 \footnotesize{ $\pm$ 0.07} \\
NeRF  & VAE & 57.326 \footnotesize{ $\pm$ 15.69} & 0.033 \footnotesize{ $\pm$ 0.02} & 0.786 \footnotesize{ $\pm$ 0.07} \\
\midrule
SDF  & AE & 94.332 \footnotesize{ $\pm$ 6.69} & 0.011 \footnotesize{ $\pm$ 0.01} & 0.847 \footnotesize{ $\pm$ 0.06} \\
SDF  & VAE & 92.497 \footnotesize{ $\pm$ 7.05} & 0.013 \footnotesize{ $\pm$ 0.00} & 0.842 \footnotesize{ $\pm$ 0.06} \\
SDF  & VQVAE & 94.355 \footnotesize{ $\pm$ 6.37} & 0.011 \footnotesize{ $\pm$ 0.01} & 0.84 \footnotesize{ $\pm$ 0.06} \\
\midrule
Shape2VecSet  & AE & 85.127 \footnotesize{ $\pm$ 13.81} & 0.019 \footnotesize{ $\pm$ 0.01} & 0.812 \footnotesize{ $\pm$ 0.06} \\
\midrule
Triplane  & AE & 63.666 \footnotesize{ $\pm$ 17.73} & 0.033 \footnotesize{ $\pm$ 0.03} & 0.761 \footnotesize{ $\pm$ 0.08} \\
Triplane  & VAE & 43.952 \footnotesize{ $\pm$ 14.02} & 0.052 \footnotesize{ $\pm$ 0.04} & 0.733 \footnotesize{ $\pm$ 0.08} \\
Triplane  & VQVAE & 45.216 \footnotesize{ $\pm$ 14.22} & 0.046 \footnotesize{ $\pm$ 0.03} & 0.733 \footnotesize{ $\pm$ 0.08} \\
\midrule
Voxel  & AE & 90.433 \footnotesize{ $\pm$ 13.14} & 0.016 \footnotesize{ $\pm$ 0.02} & 0.821 \footnotesize{ $\pm$ 0.07} \\
Voxel  & VAE & 90.500 \footnotesize{ $\pm$ 12.93} & 0.016 \footnotesize{ $\pm$ 0.02} & 0.821 \footnotesize{ $\pm$ 0.07} \\
Voxel  & VQVAE & 90.408 \footnotesize{ $\pm$ 13.03} & 0.016 \footnotesize{ $\pm$ 0.02} & 0.821 \footnotesize{ $\pm$ 0.07} \\

\bottomrule
\end{tabular}
}
\caption{Reconstruction performance by encoder (solely ShapeNet-Chair category).}
\label{tab:ablationreconstruction}
\end{table}

\begin{figure*}
    \centering
    \includegraphics[width=\linewidth]{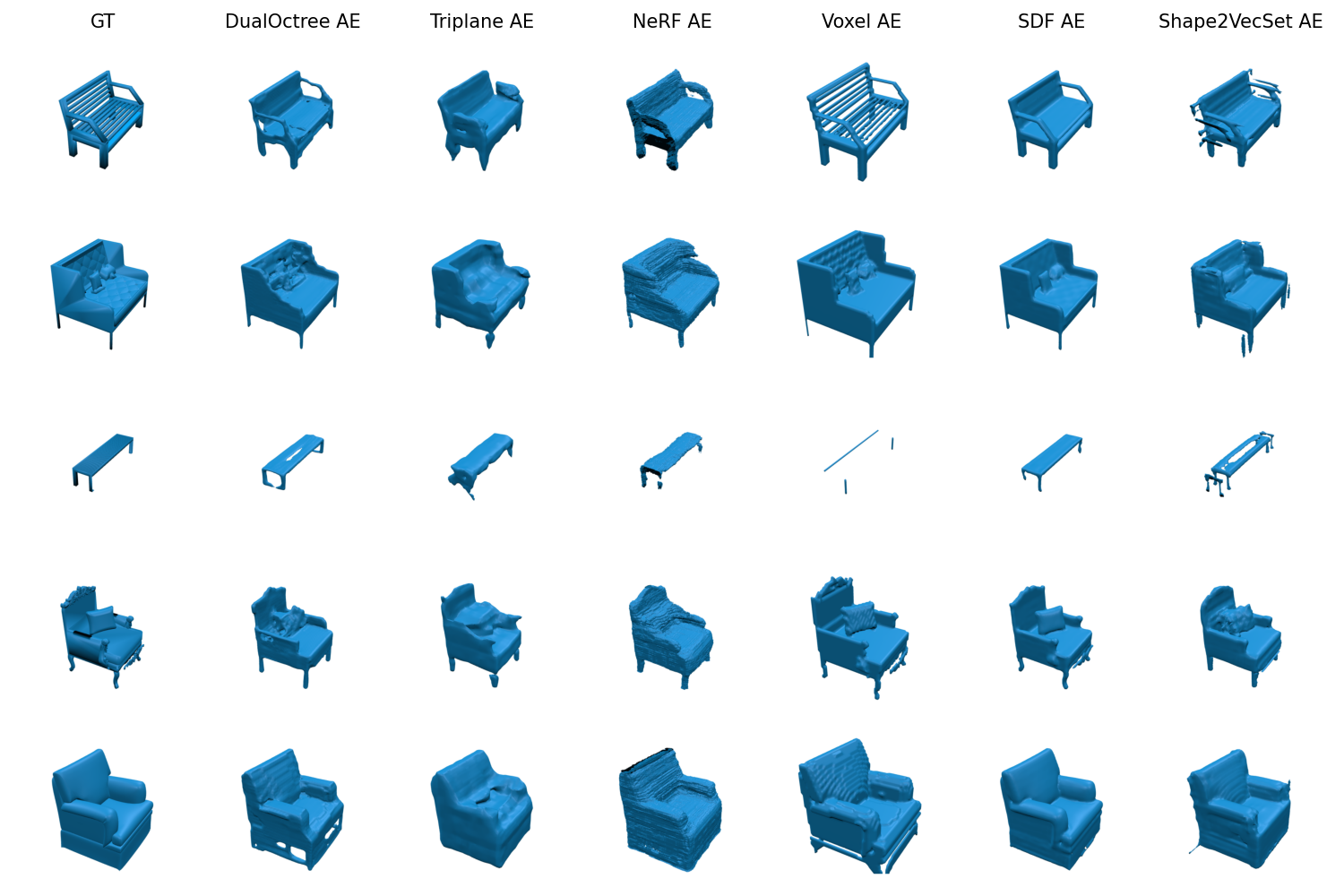}
    \caption{Qualitative results for mesh reconstruction.}
    \label{fig:reconstruction_qualitative}
\end{figure*}

Furthermore, \autoref{fig:reconstruction_qualitative} shows qualitative results for reconstruction. 
Thin or delicate structures lead to visible errors as missing parts of the object or as loss of details. This error mode is common for all methods and includes grid-less methods like NeRF.

\section{Ablation study: Generation performance}\label{app:ablation_generation}

\autoref{tab:ablation_generation} compares all variants of the generative model. The applicability of DiT and Unet depends significantly on the structure of the latent vector. In most cases, AE outperforms VQ and VQVAE. DiT is more suitable for NeRF, Voxel grid and Shape2VecSet, whereas U-Net yields better results with DualOctree, SDF grid and Triplane. Interestingly, DiT also works well for grid-based representations that were usually just used with U-Net based diffusion in related work. Due to the shorter training times and higher memory efficiency of DiT, this is a promising result for future model development.

\begin{table}[ht]
    \centering
    \resizebox{0.7\linewidth}{!}{
\begin{tabular}{lllrrr}
\toprule
\textbf{Representation} & \textbf{Generator} & \textbf{Encoder} & \textbf{COV} $\uparrow$& \textbf{MMD} $\downarrow$& \textbf{1-NNA} $\rightarrow$ 0.5\\
\midrule
DualOctree  & Unet & AE & 0.367 & 0.059 & 0.900 \\
DualOctree  & Unet & UNet & 0.458 & 0.046 & 0.686 \\
DualOctree  & Unet & VAE & 0.440 & 0.047 & 0.662 \\
DualOctree  & Unet & VQVAE & 0.393 & 0.049 & 0.708 \\
NeRF  & DiT & AE & 0.240 & 0.085 & 0.965 \\
SDF & DiT & AE & 0.432 & 0.048 & 0.726 \\
SDF  & DiT & VAE & 0.445 & 0.050 & 0.718 \\
SDF  & DiT & VQVAE & 0.385 & 0.053 & 0.728 \\
SDF  & Unet & AE & 0.372 & 0.054 & 0.787 \\
SDF  & Unet & VAE & 0.378 & 0.058 & 0.743 \\
SDF  & Unet & VQVAE & 0.347 & 0.054 & 0.744 \\
Shape2VecSet  & DiT & AE & 0.400 & 0.048 & 0.790 \\
Triplane  & Unet & AE & 0.422 & 0.052 & 0.815 \\
Voxel  & DiT & AE & 0.427 & 0.056 & 0.894 \\
Voxel  & DiT & VAE & 0.388 & 0.066 & 0.878 \\
Voxel  & DiT & VQVAE & 0.357 & 0.060 & 0.891 \\
Voxel  & Unet & AE & 0.385 & 0.059 & 0.881 \\
Voxel  & Unet & VAE & 0.380 & 0.058 & 0.828 \\
Voxel  & Unet & VQVAE & 0.438 & 0.061 & 0.826 \\
\bottomrule
\end{tabular}
    }
    \caption{Ablation study on generation performance (only ``Chair'' category). Combinations that did not converge to a state of proper 3D model generation are left out.}
    \label{tab:ablation_generation}
\end{table}

\begin{figure}
    \centering
    \includegraphics[width=0.8\linewidth]{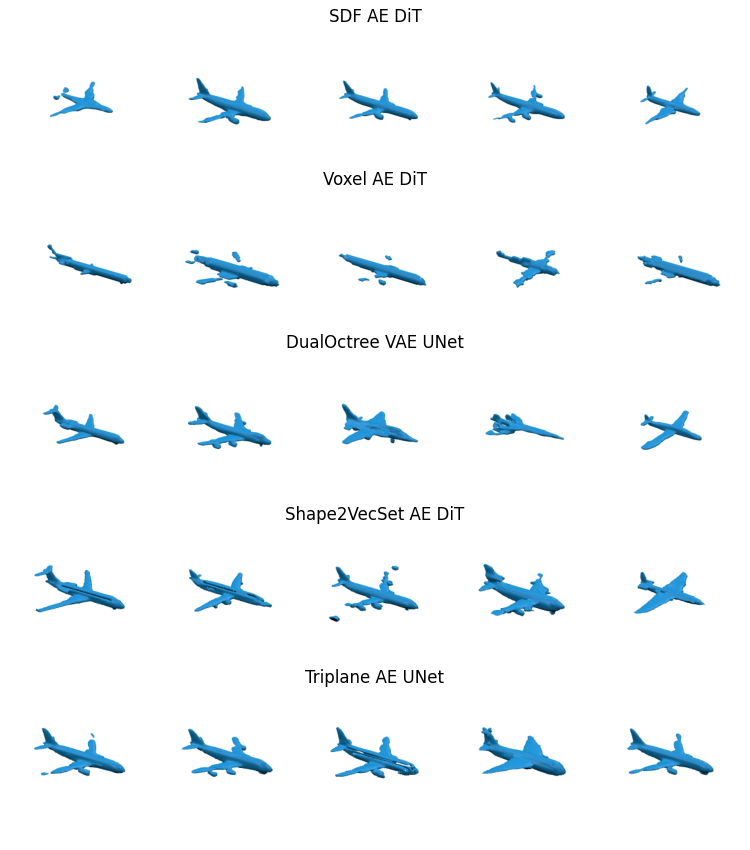}
    \caption{Qualitative results for the Airplane category}
    \label{fig:qualitative_airplane}
\end{figure}

\begin{figure}
    \centering
    \includegraphics[width=0.8\linewidth]{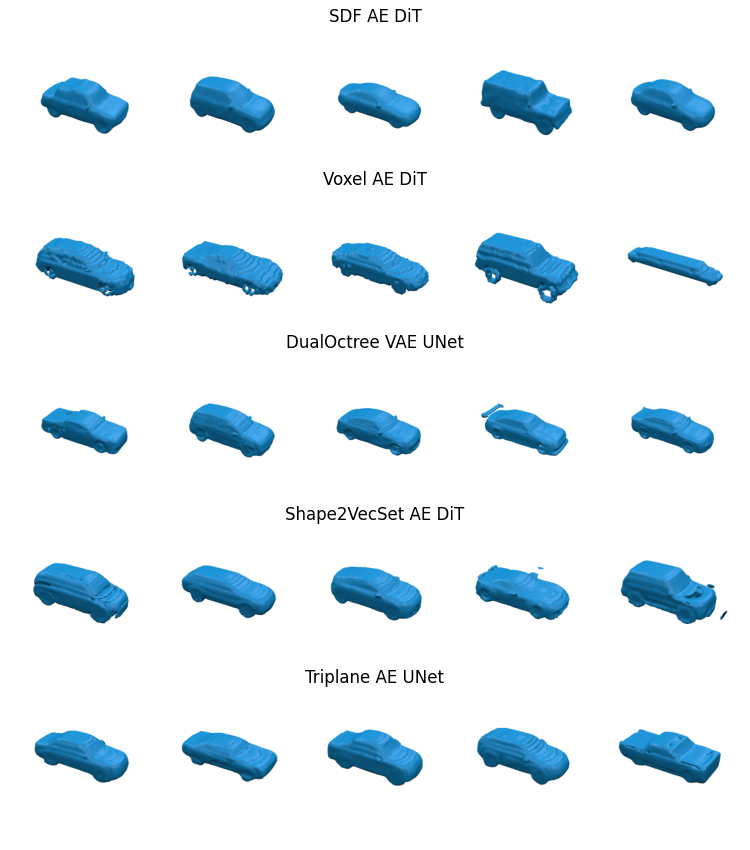}
    \caption{Qualitative results for the Car category}
    \label{fig:qualitative_car}
\end{figure}

\autoref{fig:qualitative_airplane} and \autoref{fig:qualitative_car} provide further qualitative results for the best generative approaches per representation.

\section{Mesh reconstruction under varying preprocessing methods}\label{app:airplane_recon}

\subsection{Preprocessing non-watertight meshes}

\begin{figure*}
    \centering
    \includegraphics[width=0.5\textwidth]{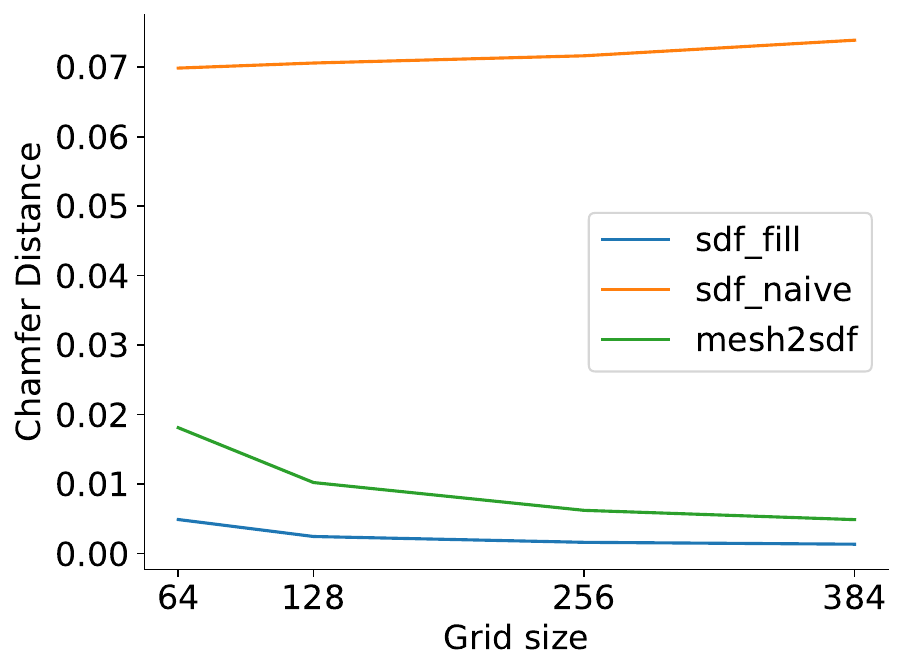}
    \caption{Comparison of different mesh conversion methods in terms of Chamfer distance.}
    \label{fig:reconstruction_error_cd}
\end{figure*}

The analysis provided in \autoref{sec:mesh_conversion} shows that some mesh preprocessing methods, such as \textit{Mesh2SDF}, substantially alter the meshes by making them thicker and thereby introduce a bias. 
In accordance with the F-score comparison in \autoref{fig:reconstruction_error_fscore}, the comparison of CD provided in \autoref{fig:reconstruction_error_cd} shows that our preprocessing step yields a mesh that diverges less from the original data.

\begin{figure}
    \centering
    \begin{subfigure}[b]{0.32\linewidth}
    \includegraphics[width=\linewidth]{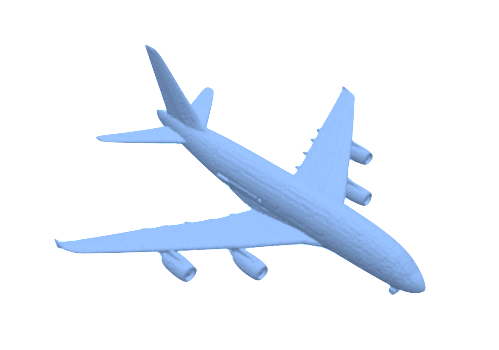}
    \caption{Ground truth}
    \end{subfigure}
    \begin{subfigure}[b]{0.32\linewidth}
    \includegraphics[width=\linewidth]{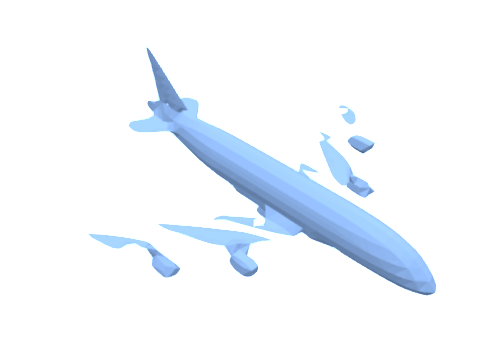}
    \caption{Reconstructed w/o \textit{Manifold}}
    \label{fig:manifold_example_wo}
    \end{subfigure}
    \begin{subfigure}[b]{0.32\linewidth}
    \includegraphics[width=\linewidth]{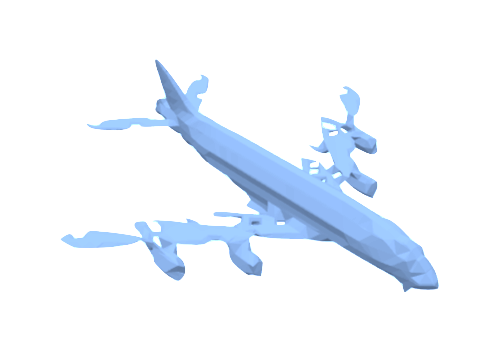}
    \caption{Reconstructed with \textit{Manifold}}
    \label{fig:manifold_example_w}
    \end{subfigure}
    \caption{Reconstruction quality when converting to a sampled SDF and back}
    \label{fig:manifold_example}
\end{figure}

\autoref{fig:manifold_example} illustrates the reconstruction of an airplane from an SDF grid with a resolution of $64^3$. We first apply the flood-fill algorithm to ensure watertightness. Then, we either directly transform the mesh into a sampled SDF and back to a mesh (\autoref{fig:manifold_example_wo}) or apply \textit{Manifold} and then transform and reconstruct the mesh (\autoref{fig:manifold_example_w}).

Without applying \textit{Manifold}, the limited grid resolution fails to capture thin structures such as airplane wings accurately, leading to incomplete or distorted reconstructions. Conversely, manifoldizing the mesh ensures that thin structures are represented as solid volumes, allowing grid-based methods to capture these features within the constraints of the grid resolution. However, this comes at the expense of altering the original mesh geometry, which may not be desirable in applications requiring high fidelity.

\subsection{Dataset quality comparison}

\begin{figure}[htb]
    \centering
    \begin{subfigure}[b]{0.45\textwidth}
        \includegraphics[width=\linewidth]{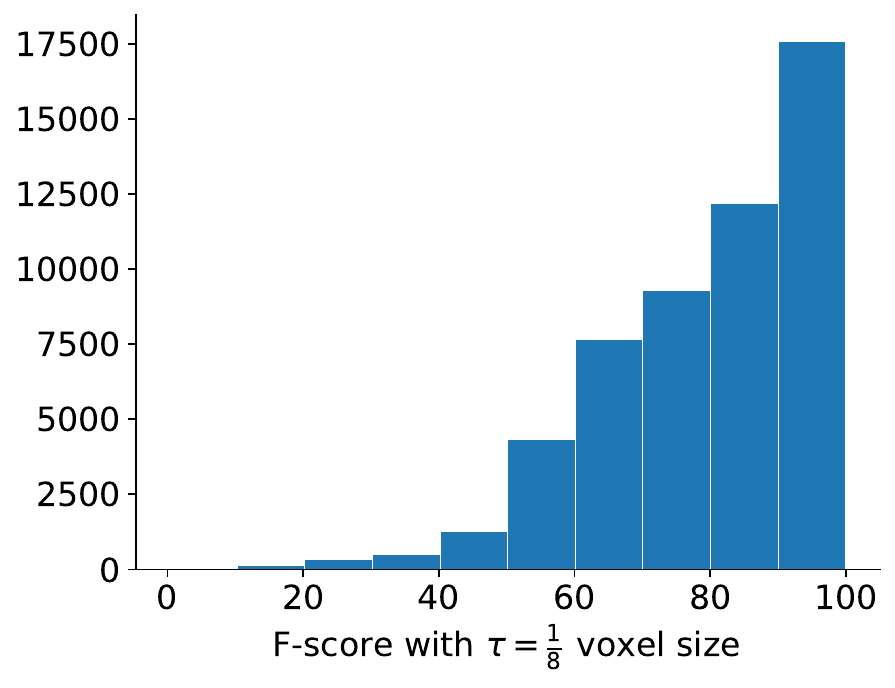}
        \caption{ShapeNet}
    \end{subfigure}\hfill
    \begin{subfigure}[b]{0.45\textwidth}
        \includegraphics[width=\linewidth]{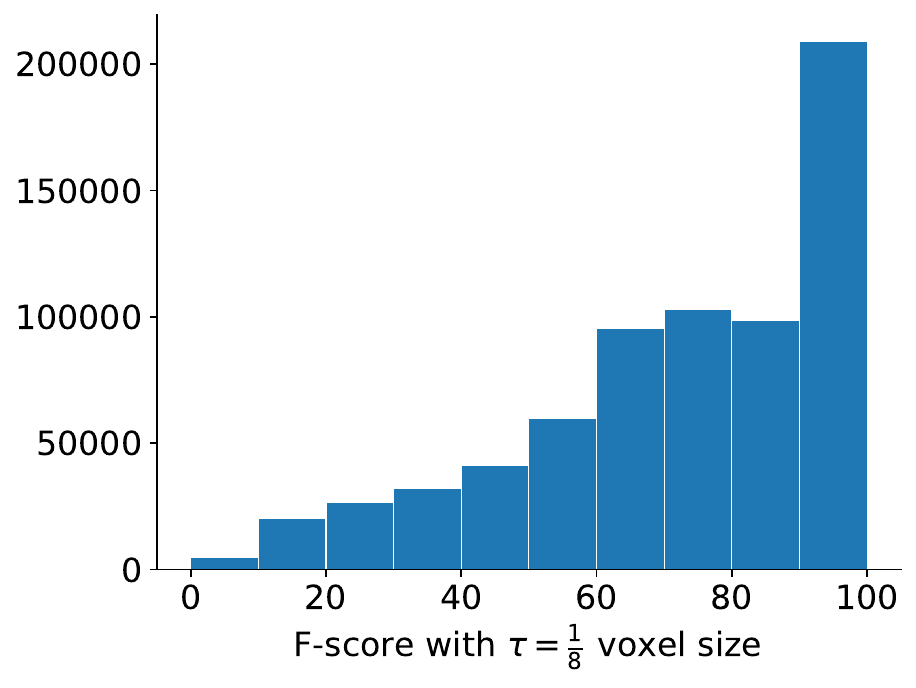}
        \caption{Objaverse}
    \end{subfigure}
    \caption{Distribution of F-scores between original and reconstructed mesh using an SDF grid of size $384^3$. The threshold for the F-score is $\tfrac{1}{8}$ of the size of a voxel.}
    \label{fig:quality_datasets}
\end{figure}

This reconstruction analysis can not only be used to compare preprocessing \textit{methods}, but also to compare the quality of \textit{datasets}. 
We analyze ShapeNet and Objaverse with our pipeline by converting all meshes to SDF grids and then convert this representation back to the mesh format. \autoref{fig:quality_datasets} shows the distribution of reconstruction errors measured here as F-score. 
We observe that a larger fraction of objects in Objaverse is of lower quality, i.e., a conversion to an implicit representation produces a larger error.

\begin{figure}[htb]
    \centering
        \includegraphics[width=0.5\linewidth]{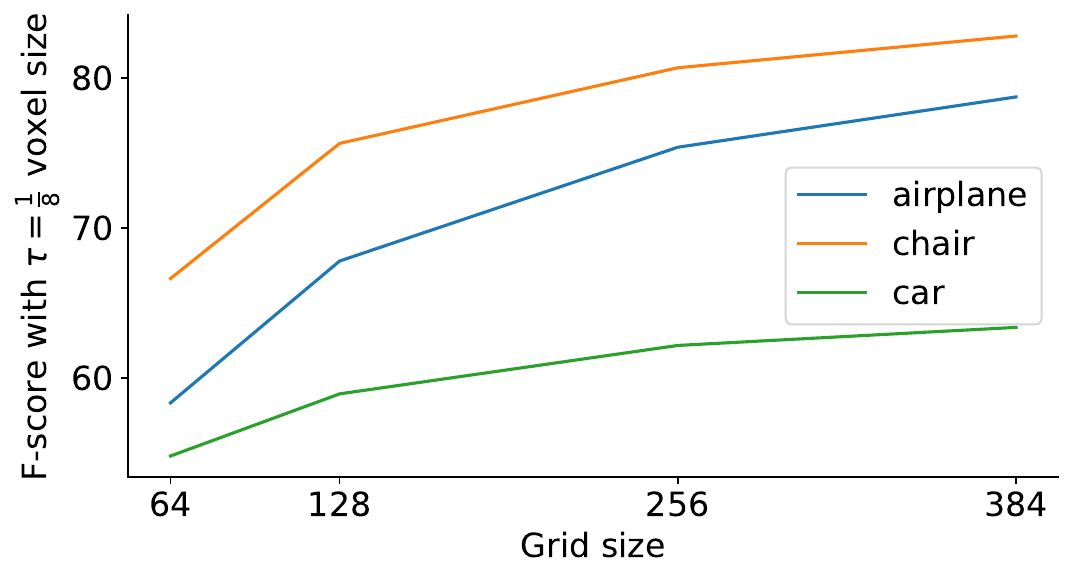}
    \caption{Effect of the grid resolution on the round trip conversion errors from mesh to SDF grid and back with our flood-fill method. Note that the threshold $\tau$ for computing the F-score is relative to the voxel size of the used grid.}
    \label{fig:quality_categories_fscpre}
\end{figure}

Further, we can observe on the ShapeNet dataset that quality differs significantly between categories. \autoref{fig:quality_categories_fscpre} shows the reconstruction error for the three categories \emph{airplane, car, chair}. The F-scores for the \emph{car} category are low even for a grid of size $384^3$, which we attribute to internal structures not captured by our preprocessing method.

\section{Representation Scalability in Complex Objects}\label{app:complexity}

We analyze how well representations handle complex shapes. We approached this in three ways: (1) We measure object complexity in the ShapeNet test set and relate it to the reconstruction performance, (2) We use shapes with sharp edges (cube), smooth surfaces (sphere), and complex fractal surfaces (Mandelbulb), and trained the methods to encode and reconstruct them, and (3) we evaluate the complexity of the \textit{generated} objects to study difficulties in generating complex objects. ``Complexity'' in (1) and (3) is measured using triangle count and surface-to-volume ratio ($\tfrac{S}{V}$).

\begin{figure}[bht]
    \centering
    \includegraphics[width=\linewidth]{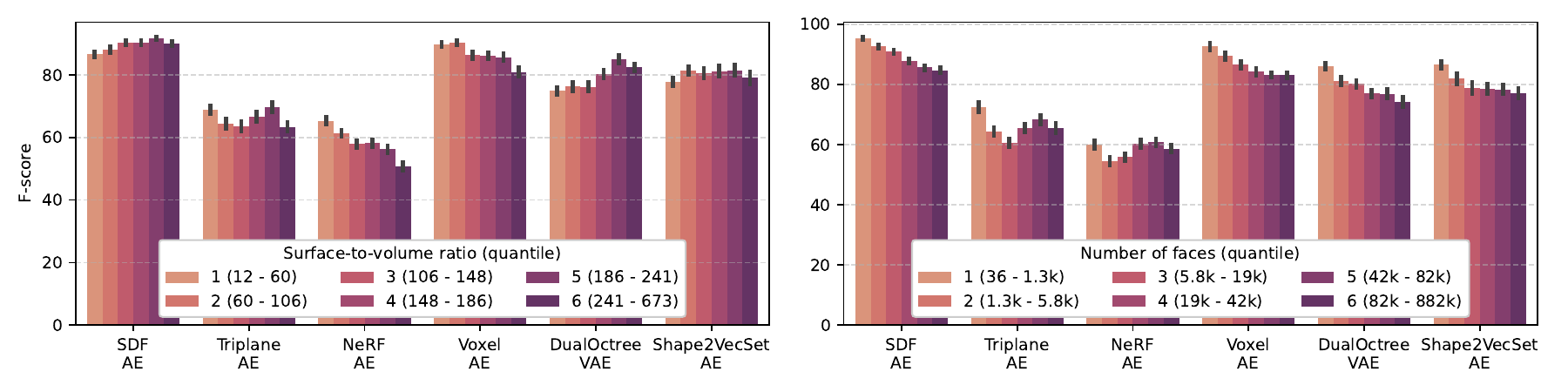}
    \caption{Complexity analysis. Best viewed zoomed in.}
    \label{fig:complexity}
\end{figure}

\begin{figure}[htb]
    \centering
    \includegraphics[width=\linewidth]{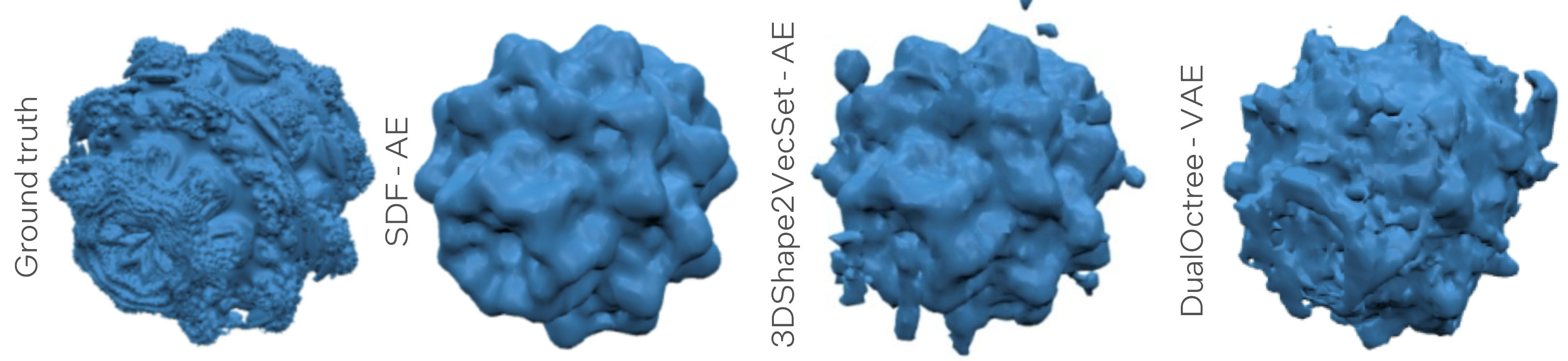}
    \caption{Mandelbulb reconstruction}
    \label{fig:mandelbulb}
\end{figure}

For approach (1), i.e. measuring object complexity in ShapeNet, \autoref{fig:complexity} (a) shows that voxel grids and NeRF suffer from higher $\tfrac{S}{V}$, whereas other representations are not affected. 
\autoref{fig:complexity} (b) shows that a higher triangle count impedes the reconstruction performance for all representations.
Shape2VecSet, however, is expected to better deal with high $\tfrac{S}{V}$ when more points are sampled. 
Furthermore, our analysis of specific complex shapes (approach (2)) shows that there are failure cases when reconstructing fractal shapes. SDF-based generation smoothes the surfaces, whereas 3DShape2VecSet introduces small artifacts (see \autoref{fig:mandelbulb}). 

\begin{figure}[b!h]
\centering
\includegraphics[width=0.7\linewidth]{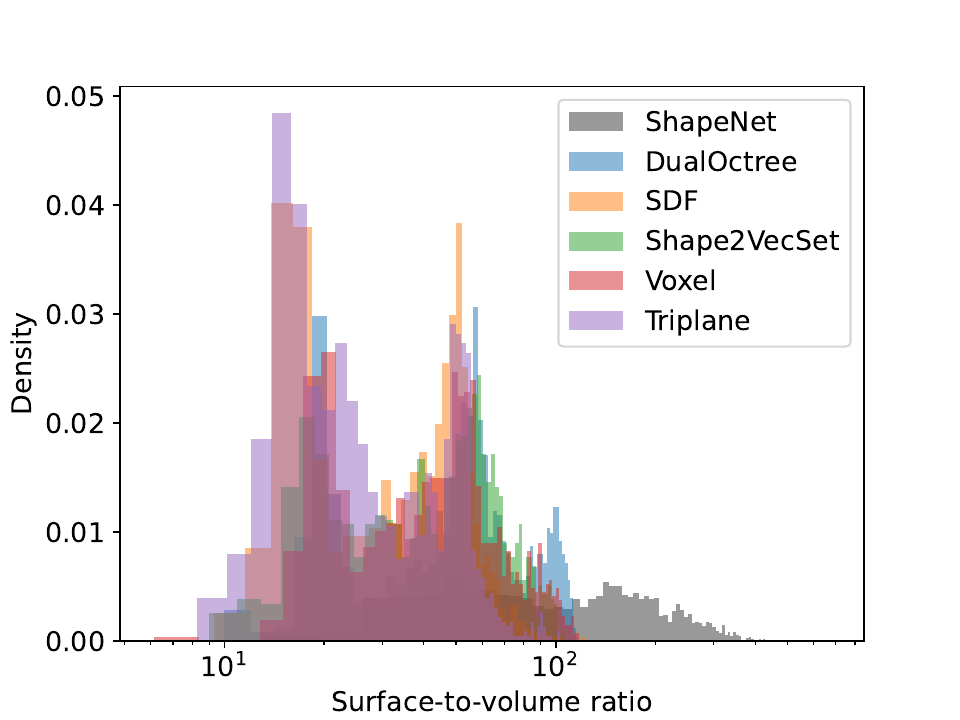}
\caption{Surface-to-volume ratio for generated objects and ShapeNet, using categories \textit{airplane}, \textit{car} and \textit{chair}.}
\label{fig:surface_volume_ratio_generated}
\end{figure}

Approach (3), i.e. evaluating the complexity of generated objects, finds that the $\tfrac{S}{V}$ is generally lower for generated objects (see \autoref{fig:surface_volume_ratio_generated}), specifically $40\pm 18$ for SDF-generated, $61\pm 27$ for DualOctree-generated, vs $156 \pm 90$ for ShapeNet objects ($N=400$). 
This suggests a bias towards generating simple, smooth objects. Training on more complex assets, e.g. Objaverse, could alleviate this issue.

\section{User Study}
\label{app:user_study}
To capture human preferences of objects generated by different methods, we asked 24 users to rate objects generated for the chair category. Each user was shown pairs of objects and asked to select the one they preferred considering the object complexity and surface quality.
To rank the methods we use the Bradley-Terry model, which models the probability of a method $A$ being better than a method $B$ as
\begin{equation}
    P(A > B) = \frac{\exp(p_A)}{\exp(p_A) + \exp(p_B)}.
\end{equation}
The parameters $p_A$ and $p_B$ are scores for the respective methods.
We estimate the scores $p$ by minimizing the cross entropy $CE$
\begin{equation}
    \min_\mathbf{p} \sum_i CE\left(\mathbf{p}, y_i)\right)
\end{equation}
with $\mathbf p$ as the vector of scores and $y_i$ as the preference labels collected from the participants.
The results are visualized in \autoref{fig:spidergraph} showing that objects generated with the SDF representations are preferred over the other representations.

\end{document}